\documentclass[10pt,journal,a4paper,twoside,twocolumn]{IEEEtran}
\usepackage{cite}
\usepackage{float}
\usepackage{stfloats}
\usepackage{fancyhdr}
\usepackage{color}
\usepackage[inline]{enumitem}
\usepackage{amsmath,amsfonts,amssymb}
\usepackage{graphicx}

\usepackage{algorithm}
\usepackage{algorithmic}
\usepackage{epsfig}
\usepackage{epstopdf}
\usepackage[caption=false]{subfig}
\usepackage{bm}
\usepackage{array}
\usepackage{bbding}
\usepackage{booktabs} 
\usepackage{multirow} 
\usepackage{pifont}
\newcommand{\cmark}{\ding{51}}

\usepackage{bbm}

\newtheorem{Rem}{Remark}

\begin{document}
	\title{DBU-OFDM: A Trainable Deep Block-Unitary OFDM Waveform for Integrated Sensing and Communication
	}
	
	\author{\IEEEauthorblockN{
		Cheng Luo, \IEEEmembership{Member, IEEE}, Luping Xiang, \IEEEmembership{Senior Member, IEEE}, Hankun Zhang, Yi Luo, Chen Huang, \IEEEmembership{Member, IEEE}, Yi Zhang, Cheng-Xiang Wang, \IEEEmembership{Fellow, IEEE} and Kun Yang, \IEEEmembership{Fellow, IEEE}
		}
			\\
		\thanks{Cheng Luo, Luping Xiang and Kun Yang are with the State Key Laboratory of Novel Software Technology, Nanjing University, Nanjing, 210008, China, and also with the Institute of Intelligent Networks and Communications (NINE) and the School of Intelligent Software and Engineering, Nanjing University, Suzhou Campus, Suzhou 215163, China. (email: chengluo@nju.edu.cn; luping.xiang@nju.edu.cn; kunyang@nju.edu.cn).}
		\thanks{Hankun Zhang and Yi Luo are with the School of Information and Communication Engineering, University of Electronic Science and Technology of China, Chengdu 611731, China. (e-mail: gy2234691702@gmail.com; 202421011115@std.uestc.edu.cn).}

        \thanks{Chen Huang and Cheng-Xiang Wang are with the National Mobile Communications Research Laboratory, School of Information Science and Engineering, Southeast University, Nanjing 211189, China, and also with Purple Mountain Laboratories, Nanjing 211111, China. (e-mail: huangchen@pmlabs.com.cn; chxwang@seu.edu.cn).}
        
        \thanks{Yi Zhang is with the Purple Mountain Laboratories, Nanjing 211111, China. (e-mail:zhangyi@pmlabs.com.cn).}
		\thanks{(Corresponding author: Luping Xiang.)}
	}

	\maketitle
	
	\thispagestyle{fancy} 
	\lhead{} 
	\chead{} 
	\rhead{} 
	\lfoot{} 
	\cfoot{} 
	\rfoot{\thepage} 
	\renewcommand{\headrulewidth}{0pt} 
	\renewcommand{\footrulewidth}{0pt} 
	\pagestyle{fancy}

    \rfoot{\thepage} 

	\begin{abstract}
	Orthogonal frequency-division multiplexing (OFDM) is a dominant waveform in modern wireless systems, yet its high peak-to-average power ratio (PAPR) and limited adaptability hinder efficient support for integrated communication and sensing. This paper proposes deep block-unitary precoded OFDM (DBU-OFDM), a structure-preserving learning framework that enables trainable waveform adaptation while preserving the DFT-based signal structure, pilot/null resource protection, and compatibility with low-complexity frequency-domain equalization. The proposed design restricts learning to a block-unitary transformation over data subcarriers and preserves pilot and null resources for structural compatibility. The transform is parameterized by recursive Householder reflections, ensuring strict unitarity as well as differentiable, numerically stable, and complexity-controllable implementation. Results show that DBU-OFDM achieves PAPR tails close to block-pilot DFT-s-OFDM while retaining comb-type pilots, improves communication reliability in frequency-selective fading via frequency-domain diversity, and enhances range and velocity estimation in direct sensing, especially in dimension-limited settings. Over-the-air USRP experiments and FPGA prototyping further verify its practical feasibility, demonstrating low error vector magnitude (EVM), clear PAPR reduction in real transmission, and hardware throughput up to 200~MS/s with microsecond-level latency. DBU-OFDM therefore offers a practical intermediate solution between conventional model-based OFDM waveforms and unconstrained neural transceivers for next-generation integrated communication and sensing systems.
	\end{abstract}

	\begin{IEEEkeywords}
		Orthogonal frequency-division multiplexing (OFDM), peak-to-average power ratio (PAPR), integrated sensing and communication (ISAC), deep learning, unitary precoding.
	\end{IEEEkeywords}

	\section{Introduction}\label{sec:I}

\IEEEPARstart{T}{he} successful deployment and widespread adoption of fifth-generation (5G) networks have accelerated the evolution toward sixth-generation (6G) wireless systems. These future networks are expected to support not only higher data rates and lower latency, but also a much tighter integration of communications, sensing, and positioning for emerging applications such as autonomous driving, immersive interaction, and intelligent manufacturing \cite{wang2023road}. In parallel with this trend, artificial intelligence (AI), particularly deep learning (DL), has emerged as a powerful tool for wireless communication innovation, offering data-driven alternatives to conventional model-based designs in increasingly complex and dynamic wireless environments \cite{kato2020ten, qin2019deep, jiang2024large,wang2025enhanced}. Within this context, waveform design is of particular importance because it directly affects spectral efficiency, hardware efficiency, receiver compatibility, and sensing capability.

From the communication perspective, recent DL-based studies have demonstrated the potential of trainable physical-layer design through constellation shaping and labeling \cite{cammerer2020trainable,o2017introduction}, end-to-end channel coding \cite{hu2024end, xiang2023polar}, pilot optimization \cite{mashhadi2021pruning, ait2021end}, and neural receiver architectures \cite{zhao2021deep}. In particular, learning-based multicarrier waveform design has shown promise in improving transmission-related metrics such as information rate, spectral containment, and peak-to-average power ratio (PAPR) \cite{ait2022waveform, huleihel2024low}, while also mitigating inter-carrier interference under strict spectral-emission constraints \cite{liu2025end}. From the sensing perspective, orthogonal frequency-division multiplexing (OFDM) and related multicarrier waveforms have also been extensively studied as dual-functional signaling frameworks, since delay, Doppler, and range parameters can be extracted directly from the communication waveform. Traditional model-based studies have developed multiple-input multiple-output (MIMO)-OFDM integrated sensing and communication (ISAC) waveforms that explicitly balance sensing and communication objectives \cite{wei2023waveform, he2024dual, zhang2024cross}. More recently, learning-based ISAC architectures have further extended waveform adaptation to highly nonconvex scenarios with practical hardware impairments and multi-target environments \cite{qi2024deep,jiang2024isac,mateos2025model}. A summary of representative prior works is provided in Table \ref{tab:comparison}.

\begin{table*}[t]
\centering
\caption{Comparison with Representative Prior Works on Learning-Based Waveform Design and ISAC}
\label{tab:comparison}
\begin{tabular}{m{5 cm}<{\centering} m{1.5 cm}<{\centering} m{1 cm}<{\centering} m{1.5 cm}<{\centering} m{1.5 cm}<{\centering} m{2 cm}<{\centering} m{1.5 cm}<{\centering}}
\toprule
\textbf{Representative Prior Works} & \textbf{Learning-based} & \textbf{Comm.} & \textbf{Sensing/ISAC} & \textbf{PAPR aware} & \textbf{OFDM Structure} & \textbf{USRP/FPGA Validation} \\
\midrule
AE-based learned communication systems \cite{cammerer2020trainable,o2017introduction,hu2024end,xiang2023polar} 
& \cmark & \cmark &  &  &  &  \\
DL-based pilot and receiver design \cite{mashhadi2021pruning,ait2021end,zhao2021deep} 
& \cmark & \cmark &  &  &  &  \\
Learning-based waveform / PAPR design \cite{ait2022waveform,huleihel2024low} 
& \cmark & \cmark &  & \cmark &  &  \\
Learning-based non-orthogonal multicarrier design \cite{liu2025end} 
& \cmark & \cmark &  &  &  &  \\
Model-based OFDM-ISAC waveform design \cite{wei2023waveform,he2024dual,zhang2024cross} 
&  & \cmark & \cmark &  & \cmark &  \\
Theoretical OFDM sensing analysis \cite{liu2025cp} 
&  &  & \cmark &  & \cmark &  \\
Learning-based ISAC designs \cite{qi2024deep,jiang2024isac,mateos2025model} 
& \cmark & \cmark & \cmark &  &  &  \\
\midrule
\textbf{Proposed DBU-OFDM} 
& \textbf{\cmark} & \textbf{\cmark} & \textbf{\cmark} & \textbf{\cmark} & \textbf{\cmark} & \textbf{\cmark} \\
\bottomrule
\end{tabular}
\vspace{-10pt}
\end{table*}

Taken together, these developments point to the need for a practical waveform framework that can jointly support communication and sensing under real-world transceiver constraints\cite{wang2025modeling}. In this regard, conventional OFDM remains an appealing foundation because of its mature transceiver architecture, flexible resource allocation capability, and compatibility with low-complexity DFT-based processing. However, precisely these structural advantages also make OFDM waveform enhancement a constrained design problem rather than a fully unconstrained learning task.

This challenge is not adequately addressed by existing AI-native designs. Many learning-based schemes remain highly task-specific, and their performance often degrades when deployment conditions differ from those seen during training \cite{akrout2024multilayer}. Moreover, the gains offered by black-box neural architectures are often difficult to interpret, which raises concerns regarding reliability, robustness, and practical integration. More fundamentally, practical OFDM waveform design must preserve the signal structure and implementation properties that enable structured resource allocation and low-complexity receiver processing. Therefore, the key problem is how to introduce trainable waveform adaptation without destroying the structural properties that make OFDM practically deployable. Under this perspective, neither rigid handcrafted designs nor fully unconstrained black-box neural transceivers provide a fully satisfactory solution.

Motivated by this observation, this paper proposes a deep block-unitary precoded OFDM waveform, termed DBU-OFDM, as a structure-preserving learning framework for OFDM waveform adaptation. Instead of learning an unconstrained transform across all subcarriers, the proposed design confines learning to a block-unitary transformation over the data subcarriers, while keeping pilot and null resources explicitly protected. By embedding the learned mapping into the DFT-based OFDM signal chain, DBU-OFDM preserves compatibility with conventional low-complexity frequency-domain equalization while enabling data-driven waveform shaping for both communication and sensing objectives. Furthermore, the trainable transform is parameterized via recursive Householder reflections, which guarantee strict unitarity and provide a differentiable, numerically stable, and complexity-controllable implementation. In this way, DBU-OFDM offers a practically viable middle ground between rigid model-based waveform design and fully unconstrained AI-native transceivers. Extensive evaluations, including over-the-air validation and FPGA-based hardware realization, demonstrate that the proposed framework can effectively enhance waveform behavior under practical structural constraints.

The main contributions of this paper are summarized as follows.
\begin{itemize}
    \item \textbf{OFDM-compatible structured waveform learning:} We identify the key requirements for practical OFDM waveform enhancement, including preservation of the cyclic-convolution and DFT-diagonalization structure, strict unitarity, and structured isolation among data, pilot, and null resource elements (REs). Based on these requirements, we propose DBU-OFDM, a structure-preserving trainable waveform architecture for OFDM systems.

    \item \textbf{Trainable unitary parameterization via Householder reflections:} We develop a structured unitary mapping for DBU-OFDM based on recursive Householder reflections. This design guarantees exact unitarity throughout training while enabling a differentiable, numerically stable, and complexity-controllable implementation for end-to-end waveform optimization.

    \item \textbf{Joint improvement in PAPR, communication, and sensing:} We show that DBU-OFDM improves waveform performance from multiple perspectives. Specifically, it achieves PAPR tail behavior close to block-pilot DFT-s-OFDM while retaining comb-type pilots, improves communication reliability in frequency-selective fading through frequency-domain diversity, and enhances direct range and velocity estimation performance, particularly when the sensing dimension is limited.

    \item \textbf{Practical validation through USRP and FPGA prototyping:} We verify the deployability of DBU-OFDM through both over-the-air experiments and hardware implementation. The results show that the proposed waveform remains compatible with the conventional OFDM processing chain, preserves nearly unchanged communication quality, achieves clear real-transmission PAPR reduction, and supports hardware throughput up to 200~MS/s with microsecond-level latency.

\end{itemize}

The rest of this paper is organized as follows. Section \ref{sec:II} presents the proposed DBU-OFDM framework, including the conventional OFDM structure and design constraints, the architecture of DBU-OFDM, and the trainable unitary matrix design. Section \ref{sec:III} evaluates the resulting communication and sensing enhancements, including PAPR reduction, communication reliability in frequency-selective fading channels, and direct sensing capability. Section \ref{sec:V} provides practical validation through over-the-air USRP experiments and FPGA implementation. Finally, Section \ref{sec:conclusion} concludes this paper.

\emph{Notation:} The notations $[\cdot]_i$ and $[\cdot]_{i,j}$ denote the $i$-th entry of a vector and the $(i,j)$-th entry of a matrix, respectively. The imaginary unit is denoted by $j \triangleq \sqrt{-1}$. The Euclidean norm and the absolute value are denoted by $\|\cdot\|$ and $|\cdot|$, respectively. The operators $(\cdot)^T$, $(\cdot)^\dagger$, and $(\cdot)^H$ denote the transpose, complex conjugate, and conjugate transpose, respectively. Finally, $\mathcal{CN}$ denotes the circularly symmetric complex Gaussian distribution.

\section{DBU-OFDM Framework}\label{sec:II}

In this paper, we investigate a DL-enhanced OFDM waveform and its practical validation. We begin by examining the key characteristics of widely adopted OFDM and summarize the fundamental constraints that should be preserved in waveform design. Based on these considerations, we then propose DBU-OFDM, a trainable OFDM-based waveform developed under such constraints, which introduces additional degrees of freedom via DL for communication and sensing. To further assess its practical viability, we develop a software-defined radio platform based on USRP and an FPGA-based prototype for hardware validation.
\begin{figure}
	\centering
	\subfloat[]{
		\includegraphics[width=0.48\linewidth]{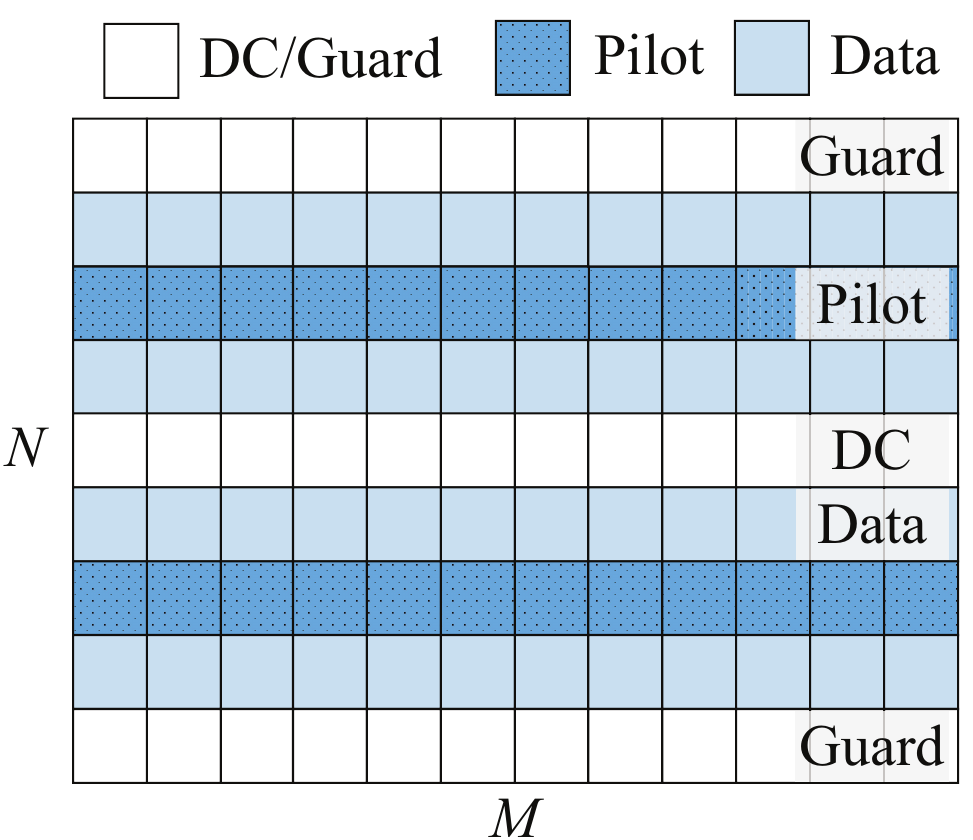}\label{fig:combtype}
	}
	\subfloat[]{
		\includegraphics[width=0.48\linewidth]{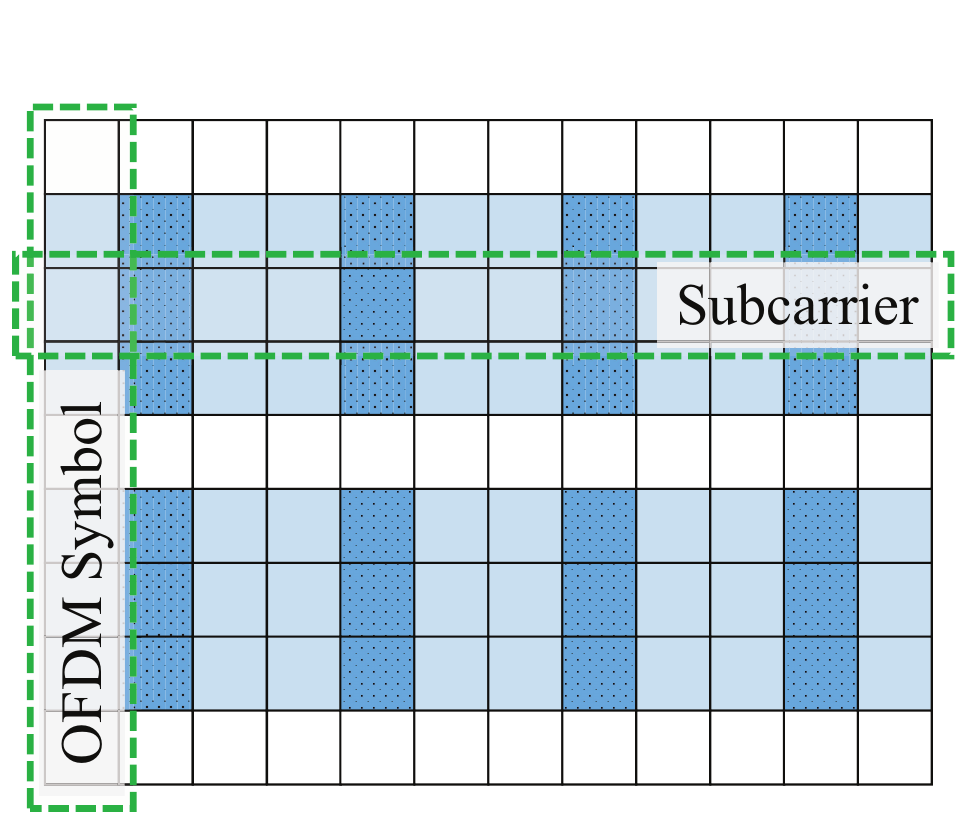}\label{fig:blocktype}
	}
	\caption{Two pilot insertion patterns in OFDM resource allocation. (a) Comb-type pilot insertion, where pilots are periodically inserted across subcarriers within each OFDM symbol. (b) Block-type pilot insertion, where pilots are clustered in dedicated OFDM symbols. }
    \vspace{-10pt}
\end{figure}

\subsection{Conventional OFDM Structure and Design Constraints}\label{sec:IIa}

We first consider a conventional OFDM system, where each frame contains $M$ consecutive OFDM symbols within one coherent processing interval (CPI), and each OFDM symbol occupies $N$ subcarriers. Among these $N$ subcarriers, $2\times N_g$ edge subcarriers are reserved as nulled guard subcarriers, $N_{dc}$ center subcarriers around the direct-current (DC) component are nulled, and $N_p$ pilot subcarriers are inserted in a comb-type pattern for channel estimation, as shown in Fig. \ref{fig:combtype}. Accordingly, the number of data subcarriers is given by
\begin{align}
	N_{\mathrm{data}} = N -2\times N_g - N_{dc} - N_p.
\end{align}

Let $S_{n,m}$ denote the modulation symbol carried by the time-frequency RE located on the $n$-th subcarrier of the $m$-th OFDM symbol, where $n\in N$ and $m\in M$. The symbols $S_{n,m}$ are drawn from a digital constellation, such as quadrature amplitude modulation (QAM). Let $\mathbf{s}_m = [S_{1,m}, \dots, S_{N,m}]^{T} \in \mathbb{C}^{N \times 1}$ denote the OFDM symbol vector, where the entries corresponding to guard and DC are set to zero, and only data and pilot subcarriers carry nonzero symbols. 

Moreover, let $\Delta f$ denote the subcarrier spacing and $T=1/\Delta f$ denote the useful OFDM symbol duration. The CP duration is denoted by $T_{\mathrm{cp}}$, and the total OFDM symbol duration is therefore denoted as 
\begin{align}
	T_o = T + T_{\mathrm{cp}}.
\end{align}


By uniformly sampling the useful OFDM symbol duration $T$ with sampling interval $T/N$, the continuous-time OFDM signal can be equivalently represented in discrete time. Accordingly, the transmitted signal of the $m$-th OFDM symbol can be compactly written as
\begin{align}
    \mathbf{x}_m = \mathbf{F}_N^{H}\mathbf{s}_m, \label{eqn:FNs}
\end{align}
where $\mathbf{F}_N \in \mathbb{C}^{N\times N}$ is the normalized $N$-point DFT matrix.

Assuming that the CP length is no smaller than the maximum channel delay spread without loss of generality, the linear convolution between the transmitted signal and the channel can be converted into a circular convolution after CP removal. Therefore, the received signal corresponding to the $m$-th OFDM symbol can be expressed as
\begin{align}
    \mathbf{y}_m = \mathbf{H}_{\mathrm{circ},m}\mathbf{x}_m + \mathbf{w}_m,\label{eqn:receivedOFDM}
\end{align}
where $\mathbf{H}_{\mathrm{circ},m} \in \mathbb{C}^{N \times N}$ denotes the circulant channel matrix and $\mathbf{w}_m\sim\mathcal{CN}(\mathbf{0},\sigma^2_0\mathbf{I}_N) \in \mathbb{C}^{N \times 1}$ is the additive noise vector.

Applying the $N$-point DFT to $\mathbf{y}_m$ yields
\begin{align}
    \mathbf{Y}_m
    &= \mathbf{F}_N \mathbf{y}_m
    = \underbrace{\mathbf{F}_N \mathbf{H}_{\mathrm{circ},m} \mathbf{F}_N^{H}}_{\mathbf{\Lambda}_m} \mathbf{s}_m + \mathbf{W}_m \nonumber\\
    &= \mathbf{\Lambda}_m \mathbf{s}_m + \mathbf{W}_m, \label{eqn:awgnrec}
\end{align}
where $\mathbf{W}_m=\mathbf{F}_N\mathbf{w}_m\sim\mathcal{CN}(\mathbf{0},\sigma_0^2\mathbf{I}_N)$ and $\mathbf{\Lambda}_m=\mathbf{F}_N\mathbf{H}_{\mathrm{circ},m}\mathbf{F}_N^{H}$ is a diagonal matrix. This result follows from the fact that the DFT basis forms the eigenbasis of any circulant matrix. Thus, once linear convolution is converted into circular convolution through CP insertion and removal, the channel matrix can be diagonalized by the DFT matrix. The resulting frequency-domain input-output relationship is therefore decoupled across subcarriers, which enables conventional OFDM to employ low-complexity single-tap equalization.

\begin{Rem}\label{remark:1}
The above observation also motivates a more general viewpoint for waveform construction. Let $\mathbf{U}\in\mathbb{C}^{N\times N}$ denote a waveform transformation matrix and define $\mathbf{x}_m = \mathbf{U}\mathbf{s}_m$ as the transformed signal. Thus, different waveforms can be interpreted through different choices of $\mathbf{U}$. For example, conventional OFDM corresponds to $\mathbf{U}=\mathbf{F}_N^H$, whereas single-carrier transmission corresponds to $\mathbf{U}=\mathbf{I}_N$. More generally, structured linear transforms also arise in other waveform families, such as $\mathbf{U}=\mathbf{F}_M^H\otimes\mathbf{I}_L$ for orthogonal time-frequency space (OTFS)\cite{hadani2017orthogonal}, and $\mathbf{U}=\mathbf{\Lambda}_{c_1}\mathbf{F}_N^H\mathbf{\Lambda}_{c_2}$ for affine frequency division multiplexing (AFDM)\cite{bemani2023affine}, where $\mathbf{\Lambda}_{c_i}, \forall i\in\{1,2\}$ denotes a diagonal chirp matrix parameterized by $c_i$. From this perspective, waveform design can be viewed as the design of a structured transformation matrix, which motivates the proposed DBU-OFDM design.
\end{Rem}

However, for an OFDM-enhanced waveform, transformation matrix $\mathbf{U}$ cannot be chosen arbitrarily. To preserve the key advantages of conventional OFDM and maintain compatibility with existing transceiver procedures, the waveform transformation should satisfy the following fundamental requirements.

\begin{enumerate}
    \item \textit{Preservation of the cyclic-convolution and DFT-diagonalization structure:}
    the introduced waveform processing should not destroy the circulant channel structure after CP removal, such that the equivalent channel remains diagonalizable by the DFT basis and the conventional OFDM equalization framework can still be retained.

    \item \textit{Preservation of unitary structure:}
    the transformation should remain unitary, such that signal energy is preserved and noise enhancement is avoided during modulation and recovery.

    \item \textit{Structured isolation among data, pilot, and null subcarriers/REs:}
    data, pilot, and null subcarriers, including DC and guard subcarriers, should remain controllably separated, so as to preserve channel estimation accuracy, spectral containment, and compatibility with practical OFDM receiver operations.
\end{enumerate}

These requirements define the feasible design region for OFDM-compatible waveform enhancement. Building upon them, the next section introduces the proposed DBU-OFDM architecture, which incorporates additional trainable degrees of freedom while preserving the essential structure of OFDM.

\subsection{Architecture of DBU-OFDM}\label{sec:IIb}

In this section, we first describe the architecture of the proposed DBU-OFDM waveform. We then present a practical approach for training the associated unitary transformation matrix under the imposed structural constraints.

\begin{figure}
	\centering
	\includegraphics[width=0.998\linewidth]{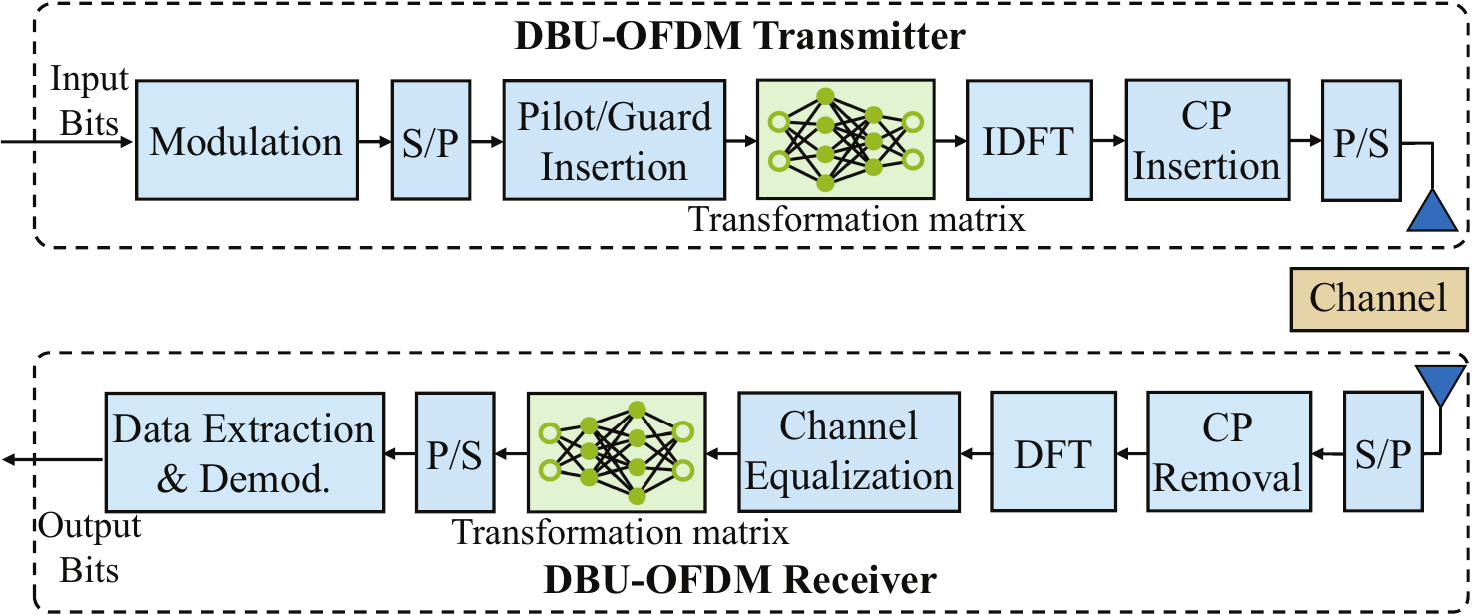}
	\caption{End-to-end transceiver pipeline of the proposed DBU-OFDM system. The blue blocks denote conventional OFDM processing modules, while the green blocks represent the proposed trainable transformation matrix.}
    \vspace{-10pt}
	\label{fig:DBUOFDMentirePipeline}
\end{figure}

Motivated by the design constraints summarized in Section~\ref{sec:IIa}, we now introduce the proposed DBU-OFDM architecture. The overall transceiver architecture is illustrated in Fig. \ref{fig:DBUOFDMentirePipeline}.

We first consider a waveform transformation matrix $\mathbf{U}\in\mathbb{C}^{N\times N}$, yielding
\begin{align}
    \bar{\mathbf{s}}_m = \mathbf{U}\mathbf{s}_m.
\end{align}
To avoid noise enhancement and preserve signal energy, $\mathbf{U}$ is first required to be unitary. However, an unconstrained unitary matrix generally cannot maintain the structured isolation among data, pilot, and null REs, and may therefore lead to undesired mixing across different RE types. 

A direct optimization of a full unitary matrix $\mathbf{U}$, followed by puncturing or masking over selected subcarriers, can enforce such isolation\cite{abrudan2022unitary}, but may easily destroy strict unitarity and introduce undesirable energy leakage across different resource types. 

To address this issue, we adopt an architecture based on \emph{functional block construction} and \emph{index mapping}. Specifically, the subcarriers are partitioned into three groups, namely data, pilot, and null subcarriers, with cardinalities $N_{\mathrm{data}}$, $N_p$, and $N_{\mathrm{null}}$, respectively, satisfying
\begin{align}
    N_{\mathrm{data}} + N_p + N_{\mathrm{null}} = N.
\end{align}

We then define a block-diagonal unitary matrix
\begin{align}
    \mathbf{U}'=\mathrm{blkdiag}\!\left(\mathbf{U}_{\mathrm{data}},\mathbf{U}_{\mathrm{pilot}},\mathbf{U}_{\mathrm{null}}\right),
    \label{eqn:blockU}
\end{align}
where
\begin{itemize}
    \item $\mathbf{U}_{\mathrm{data}}\in\mathbb{C}^{N_{\mathrm{data}}\times N_{\mathrm{data}}}$ is the trainable unitary block associated with data subcarriers. It contains the learnable degrees of freedom of the proposed waveform and is responsible for data-domain shaping and adaptation. 

    \item $\mathbf{U}_{\mathrm{pilot}}\in\mathbb{C}^{N_p\times N_p}$ is the unitary block associated with pilot subcarriers. In this work, it is fixed as $\mathbf{U}_{\mathrm{pilot}}=\mathbf{I}_{N_p}$ to preserve pilot purity and compatibility with conventional channel estimation. However, in advanced joint waveform and pilot optimization frameworks, $\mathbf{U}_{\mathrm{pilot}}$ could also be extended to a trainable block.

    \item $\mathbf{U}_{\mathrm{null}}\in\mathbb{C}^{N_{\mathrm{null}}\times N_{\mathrm{null}}}$ is the unitary block associated with null subcarriers, including DC and guard subcarriers. In this work, it is fixed as $\mathbf{U}_{\mathrm{null}}=\mathbf{I}_{N_{\mathrm{null}}}$ to enforce strict spectral protection and prevent leakage into protected frequency resources.
\end{itemize}
Note that if $\mathbf{U}_{\mathrm{data}}$, $\mathbf{U}_{\mathrm{pilot}}$, and $\mathbf{U}_{\mathrm{null}}$ are set to the identity matrix, the proposed DBU-OFDM simplifies to conventional OFDM.


\begin{figure*}
	\centering
	\includegraphics[width=0.7\linewidth]{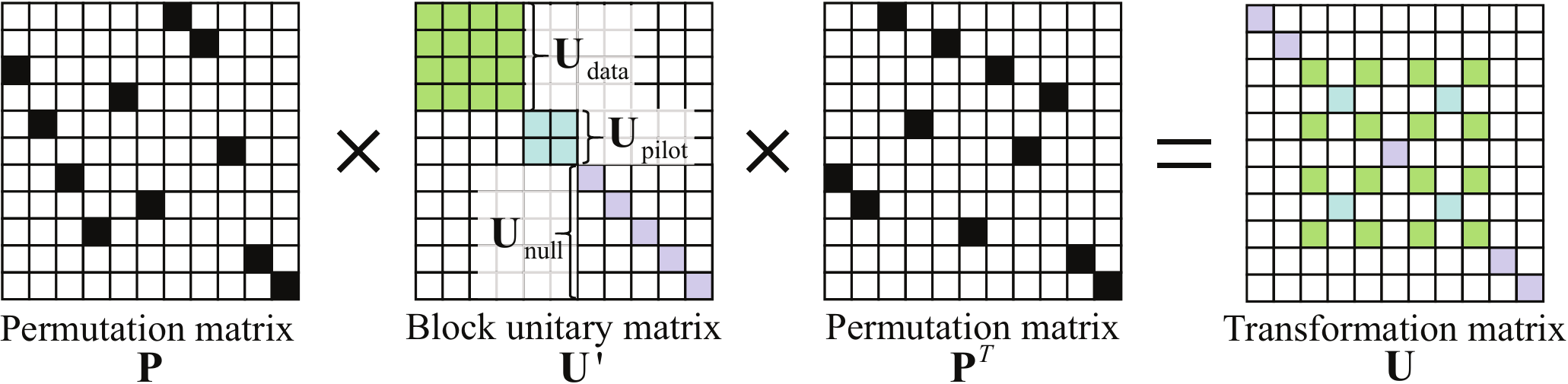}
	\caption{Construction of the DBU-OFDM transformation matrix. The permutation matrix contains binary entries and realizes structured index remapping across resource groups.}
    \vspace{-10pt}
	\label{fig:DBUofdmpipeline}
\end{figure*}

Then, a permutation matrix $\mathbf{P}\in\{0,1\}^{N\times N}$ is introduced to realize structured resource mapping, as shown in Fig.~\ref{fig:DBUofdmpipeline}. Thus, the global transformation matrix is constructed as
\begin{align}
    \mathbf{U} = \mathbf{P}\mathbf{U}'\mathbf{P}^{T}.
    \label{eqn:globalU}
\end{align}
Since each diagonal block of $\mathbf{U}'$ is unitary and $\mathbf{P}$ is an orthogonal matrix satisfying
\begin{align}
    \mathbf{P}^{T}\mathbf{P}=\mathbf{P}\mathbf{P}^{T}=\mathbf{I}_N,
\end{align}
the global matrix $\mathbf{U}$ is also unitary, i.e.,
\begin{align}
    \mathbf{U}^{H}\mathbf{U}
    &= \left(\mathbf{P}\mathbf{U}'\mathbf{P}^{T}\right)^{H}
       \left(\mathbf{P}\mathbf{U}'\mathbf{P}^{T}\right) = \mathbf{P}\left(\mathbf{U}'\right)^{H}\mathbf{P}^{T}\mathbf{P}\mathbf{U}'\mathbf{P}^{T} \nonumber\\
    &= \mathbf{P}\left(\mathbf{U}'\right)^{H}\mathbf{U}'\mathbf{P}^{T} = \mathbf{P}\mathbf{I}_N\mathbf{P}^{T} = \mathbf{I}_N.
    \label{eqn:U_unitary}
\end{align}
Thus, transformation matrix $\mathbf{U}$ preserves signal energy and avoids noise enhancement during transformation and recovery. Moreover, owing to the block structure, the transformed energy remains confined within the corresponding subcarrier groups/REs.

\begin{Rem}
The above construction has a clear physical interpretation. The trainable block $\mathbf{U}_{\mathrm{data}}$ is responsible for adaptive feature rotation over data subcarriers, thereby enabling frequency-domain shaping, PAPR reduction, and diversity enhancement. In contrast, the pilot and null blocks remain structurally protected, i.e., $\mathbf{U}_{\mathrm{pilot}}=\mathbf{I}_{N_{p}}$ and $\mathbf{U}_{\mathrm{null}}=\mathbf{I}_{N_{\mathrm{null}}}$, which prevents the learning process from contaminating channel estimation resources or violating guard-band constraints. The permutation matrix $\mathbf{P}$, typically derived from the classical OFDM resource structure (e.g., the 3GPP frame structure), further enables dynamic yet structured index remapping without introducing any additional amplitude or phase distortion.
\end{Rem}

The final constraint to be satisfied is the diagonalization of the circulant channel. Although this property may be encouraged by introducing additional loss functions, such an approach does not guarantee mathematically exact diagonalization. Therefore, we explicitly combine the transformation matrix $\mathbf{U}$ with the DFT-based OFDM structure and define the final transformation matrix $\mathbf{T}$ as
\begin{align}
    \mathbf{T} = \mathbf{F}_N^{H}\mathbf{U},
\end{align}
where $\mathbf{T}$ is also unitary since
\begin{align}
    \mathbf{T}^H\mathbf{T}
    = \mathbf{U}^H\mathbf{F}_N\mathbf{F}_N^{H}\mathbf{U}
    = \mathbf{I}_N.
\end{align}

Thus, similar to Eq.~\eqref{eqn:receivedOFDM}, with the transformation matrix $\mathbf{T}$, the received signal after CP removal can be expressed as
\begin{align}
    \mathbf{y}_m = \mathbf{H}_{\mathrm{circ},m}\underbrace{\mathbf{F}_N^{H}\mathbf{U}}_{\mathbf{T}}\mathbf{s}_m + \mathbf{w}_m.
\end{align}
Applying the DFT at the receiver yields
\begin{align}
    \mathbf{Y}_m
    &= \mathbf{F}_N\mathbf{y}_m \nonumber\\
    &= \mathbf{F}_N\mathbf{H}_{\mathrm{circ},m}\mathbf{F}_N^{H}\mathbf{U}\mathbf{s}_m + \mathbf{W}_m \nonumber\\
    &= \mathbf{\Lambda}_m \mathbf{U}\mathbf{s}_m + \mathbf{W}_m.
    \label{eqn:dbu_rx}
\end{align}
Since the insertion of $\mathbf{U}$ does not alter the CP-induced circulant channel structure, the equivalent channel remains diagonalized by the DFT basis. Therefore, the proposed DBU-OFDM waveform retains the conventional low-complexity OFDM equalization framework, while augmenting the transmit signal with additional trainable and structure-preserving waveform flexibility. After single-tap equalization, the transformed symbol vector $\mathbf{U}\mathbf{s}_m$ can be recovered, and the original symbol vector $\mathbf{s}_m$ is then obtained through the inverse transformation $\mathbf{U}^H$.

\subsection{Trainable Unitary Matrix Design}\label{sec:ii-c}

\begin{figure}
	\centering
	\includegraphics[width=0.99\linewidth]{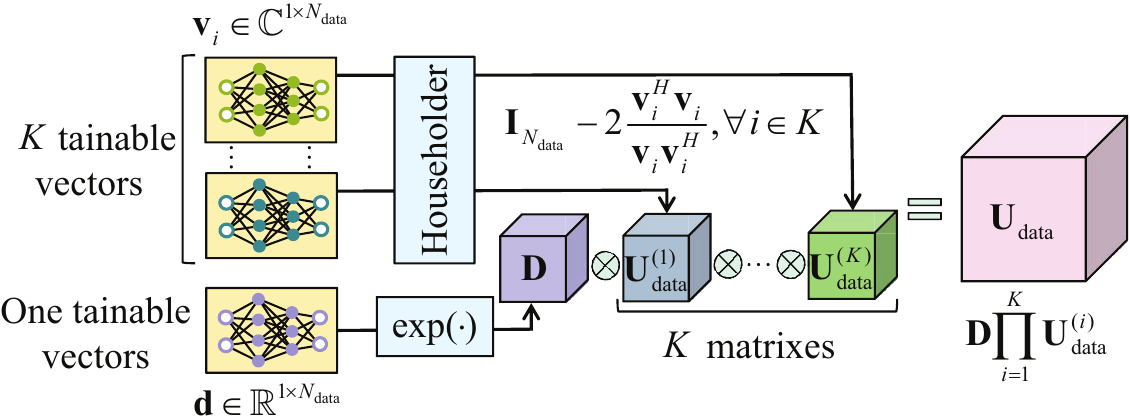}
	\caption{Recursive construction of the trainable unitary matrix via Householder reflections.}
    \vspace{-10pt}
	\label{fig:householdeTF}
\end{figure}

To guarantee strict unitarity throughout training, the trainable block $\mathbf{U}_{\mathrm{data}}$ should be parameterized through structure-preserving unitary generation methods. In this paper, we adopt a trainable unitary matrix generation method based on Householder reflections, as shown in Fig. \ref{fig:householdeTF}. By introducing trainable vectors for recursive construction, the resulting unitary matrix can be generated in a differentiable, numerically stable, and complexity-controllable manner.

Specifically, we first generate $K$ unconstrained trainable complex vectors, denoted by $\mathbf{v}_i \in \mathbb{C}^{1\times N_{\mathrm{data}}},\forall i\in K$, which are iteratively updated during training. Instead of directly optimizing a highly constrained unitary matrix, each vector $\mathbf{v}_i$ is mapped to a differentiable Householder matrix via the Householder reflection, given by
\begin{align}
\mathbf{U}_{\mathrm{data}}^{(i)}
= \mathbf{I}_{N_{\mathrm{data}}}-2 \frac{\mathbf{v}_i^H \mathbf{v}_i}{\mathbf{v}_i \mathbf{v}_i^H},
\end{align}
By construction, each $\mathbf{U}_{\mathrm{data}}^{(i)}$ is both Hermitian and unitary, while remaining fully differentiable with respect to the underlying trainable parameters.

To further enrich the phase representation capability, we additionally introduce an unconstrained real-valued trainable vector $\mathbf{d}=[d_1,\ldots,d_{N_{\mathrm{data}}}] \in \mathbb{R}^{1\times N_{\mathrm{data}}}$. Its entries are mapped through the complex exponential function to construct a diagonal phase matrix as 
\begin{align}
\mathbf{D}=
\operatorname{diag}
\left(
e^{j d_1}, e^{j d_2}, \ldots, e^{j d_{N_{\mathrm{data}}}}
\right),
\end{align}
whose diagonal entries all have unit modulus. Therefore, $\mathbf{D}$ is unitary by construction.

Subsequently, as illustrated in Fig.~\ref{fig:householdeTF}, the $K$ Householder matrices are sequentially multiplied and further combined with the diagonal phase matrix $\mathbf{D}$ to synthesize the final differentiable unitary matrix $\mathbf{U}_{\mathrm{data}} \in \mathbb{C}^{N_{\mathrm{data}} \times N_{\mathrm{data}}}$, i.e.,
\begin{align}
\mathbf{U}_{\mathrm{data}}
= \mathbf{D}\prod_{i=1}^{K}\mathbf{U}_{\mathrm{data}}^{(i)}
= \mathbf{D}\mathbf{U}_{\mathrm{data}}^{(1)}\mathbf{U}_{\mathrm{data}}^{(2)}\cdots \mathbf{U}_{\mathrm{data}}^{(K)}.
\end{align}
Since both $\mathbf{D}$ and each $\mathbf{U}_{\mathrm{data}}^{(i)}$ are unitary, their product remains unitary by construction.

Through this parameterization, the proposed DBU-OFDM architecture incorporates the unitary constraint directly into the forward transformation, thereby avoiding computationally intensive re-orthogonalization procedures, e.g., singular value decomposition (SVD), during training. Moreover, the inclusion of the diagonal phase matrix $\mathbf{D}$ provides additional phase flexibility, resulting in a more expressive, differentiable, numerically stable, and complexity-controllable implementation for end-to-end optimization.

\begin{Rem}
Note that the number of Householder factors $K$ provides a direct mechanism for controlling the complexity of the trainable unitary transformation. In particular, each additional Householder factor introduces one extra trainable reflection vector, thereby expanding the degrees of freedom and correspondingly increasing the expressiveness of the resulting matrix $\mathbf{U}_{\mathrm{data}}$. Therefore, $K$ offers a flexible complexity-performance tradeoff for practical implementation.

Moreover, when $K \geq N_{\mathrm{data}}-1$, the above parameterization, together with the diagonal phase matrix $\mathbf{D}$, is sufficiently expressive to represent an arbitrary unitary matrix in $\mathbb{C}^{N_{\mathrm{data}}\times N_{\mathrm{data}}}$. This follows from the classical Householder factorization\cite{ivanov2006engineering}: for any target unitary matrix $\tilde{\mathbf{U}} \in \mathbb{C}^{N_{\mathrm{data}}\times N_{\mathrm{data}}}$, one can successively apply at most $N_{\mathrm{data}}-1$ Householder reflections to transform its columns into the standard basis, leaving only a diagonal unit-modulus matrix. Therefore, the proposed construction is universal in the sense of unitary representation.

However, in practical systems, $K$ does not need to be chosen as large as $N_{\mathrm{data}}-1$. In many cases, a much smaller value of $K$ is sufficient to capture the dominant transformation components required by the task. From this perspective, the proposed parameterization acts as a structural bottleneck on the degrees of freedom, which can reduce implementation complexity while retaining the most relevant waveform features for end-to-end optimization.
\end{Rem}

\section{Communication and Sensing Enhancements of DBU-OFDM}\label{sec:III}

In this section, we investigate the performance enhancements enabled by the proposed DBU-OFDM architecture. The following subsections systematically evaluate its PAPR reduction capability, communication reliability in frequency-selective fading channels, and direct sensing functionality supported by the optimized waveform structure. The parameters are configured in Table \ref{tab:ofdm_parameters} unless otherwise specified.

\begin{table}[htbp]
\centering
\caption{Parameter Configurations}
\label{tab:ofdm_parameters}
\begin{tabular}{cccc}
\hline
\textbf{Parameter} & \textbf{Config. 1} & \textbf{Config. 2} & \textbf{Config. 3} \\ 
\hline
Total Subcarriers ($N$)       & 64           & 128          & 256          \\
Cyclic Prefix Length ($N_{cp}$)& 16           & 32           & 64           \\
Guard Subcarriers ($N_g$)     & $2 \times 4$ & $2 \times 8$ & $2 \times 16$\\
Pilot Subcarriers ($N_p$)     & 8            & 16           & 16           \\
DC Subcarriers ($N_{dc}$)     & \multicolumn{3}{c}{2}                    \\ 
OFDM symbols ($M$)     & \multicolumn{3}{c}{8}                    \\ 
\hline
\end{tabular}
\vspace{-10pt}
\end{table}

\subsection{PAPR Reduction}
One of the most critical practical issues in OFDM-type waveforms is the high PAPR. Owing to the coherent superposition of multiple subcarriers in the time domain, the instantaneous envelope of OFDM signals may exhibit large peaks, which in turn imposes stringent linearity requirements on the radio-frequency (RF) front-end, especially the power amplifier (PA). In practical hardware systems, a high PAPR typically forces the PA to operate with a large input back-off, resulting in reduced power efficiency and degraded hardware utilization. Therefore, in addition to average communication performance, PAPR is a key waveform metric that must be explicitly considered in practical waveform design and prototype validation.

For hardware-oriented evaluation, the complementary cumulative distribution function (CCDF) of PAPR is particularly important, since it characterizes the probability that the instantaneous PAPR exceeds a given threshold. This metric directly reflects the likelihood of rare but harmful high-power peaks, and is therefore widely adopted in practice to assess the PA-friendliness of multicarrier waveforms.

Let $\mathbf{x}_m = [x_m[1],x_m[2],\ldots,x_m[N_s]]^T$ denote the discrete-time time-domain signal of the $m$-th transmitted symbol, where $N_s$ denotes the number of samples used for PAPR evaluation.\footnote{In practical evaluation, $N_s$ may correspond to the original symbol length or an oversampled version for more accurate peak estimation.} The PAPR (dB) of $\mathbf{x}_m$ is defined as
\begin{align}
     \mathrm{PAPR}_{\mathrm{dB}}(\mathbf{x}_m)=
    10\log_{10}\!\left(\frac{\max_{1\le n\le N_s}|x_m[n]|^2}
    {\frac{1}{N_s}\sum_{n=1}^{N_s}|x_m[n]|^2}\right).
\end{align}

Accordingly, the CCDF of PAPR at a threshold $\gamma$ dB is defined as
\begin{align}
    \mathrm{CCDF}(\gamma)
    = \Pr\!\left(\mathrm{PAPR}_{\mathrm{dB}}(\mathbf{x}_m) > \gamma \right).
\end{align}
Although the CCDF is the standard metric for PAPR evaluation, it is not directly suitable as a training objective, since the corresponding indicator function is non-differentiable. To enable end-to-end optimization, we therefore introduce a differentiable proxy loss inspired by the CCDF criterion.

Specifically, let $\gamma_{\mathrm{tar}}$ denote the target PAPR threshold in dB. For each transmitted symbol, we define the exceedance error as
\begin{align}
    e_m = \operatorname{ReLU}\!\left(\mathrm{PAPR}_{\mathrm{dB}}(\mathbf{x}_m)-\gamma_{\mathrm{tar}}\right),
\end{align}
which penalizes only those samples whose PAPR exceeds the target threshold, while ignoring symbols already below the desired level. Based on this exceedance term, the proposed CCDF-inspired proxy loss is defined as
\begin{align}
    \mathcal{L}_{\mathrm{PAPR}}
    = \frac{1}{M}\sum_{m=1}^{M} e_m^{\,p},
    \label{eq:papr_loss}
\end{align}
where $M$ denotes the number of training samples in the batch, and $p\in\{1,2\}$ is a hardness factor controlling the penalty strength. In particular, $p=1$ corresponds to a linear penalty, while $p=2$ imposes a stronger quadratic penalty on symbols with excessive PAPR.

The above loss can be viewed as a differentiable approximation to CCDF minimization. Instead of explicitly counting the proportion of samples above a threshold, it penalizes the amount by which the PAPR exceeds the desired target. As a result, the loss decreases when fewer samples violate the threshold and when the exceedance of violating samples becomes smaller, which is fully consistent with the physical objective of reducing the PAPR CCDF.

\begin{figure}
	\centering
	\subfloat[]{
		\includegraphics[width=0.8\linewidth]{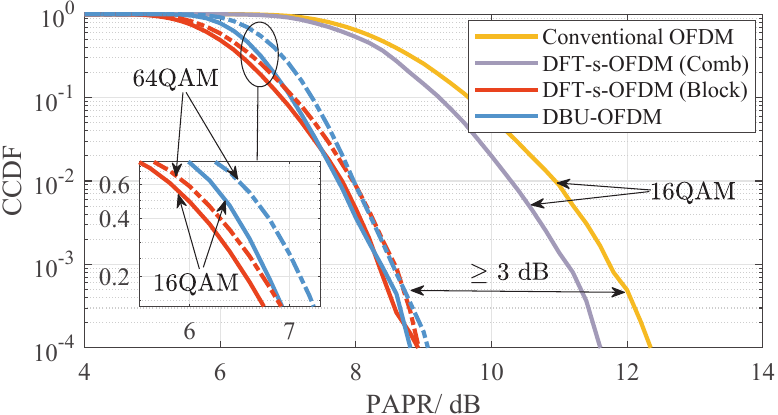}\label{fig:PAPR_org}
	}\\
	\vspace{-10pt}
	\subfloat[]{
		\includegraphics[width=0.8\linewidth]{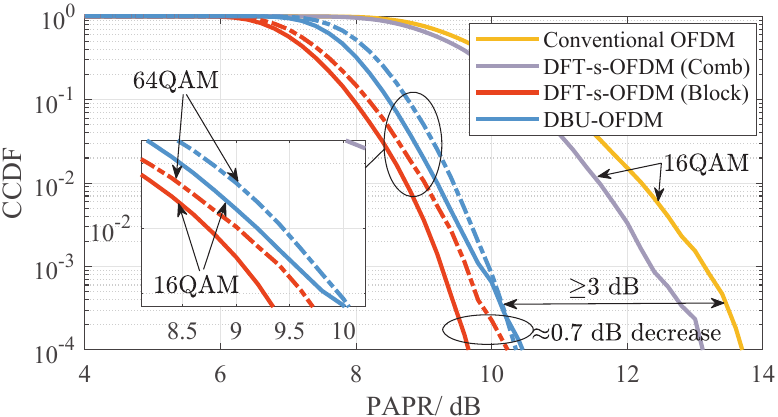}\label{fig:PAPR_4times}
	}
	\caption{CCDFs for different waveforms, with parameters $N=256$, $N_g=16$, $N_{dc}=2$, $N_p=16$ and $N_{cp}=64$.}
    \vspace{-10pt}
    \label{fig:PAPR_all}
\end{figure}

We evaluate the PAPR performance of conventional OFDM, DFT-spread OFDM (DFT-s-OFDM), and the proposed DBU-OFDM. Conventional OFDM and DBU-OFDM employ the comb-type pilot pattern shown in Fig. \ref{fig:combtype}, while DFT-s-OFDM is evaluated with both block-type and comb-type pilot configurations. The corresponding CCDF curves of the PAPR are shown in Fig. \ref{fig:PAPR_org}.

With block-type pilots, DFT-s-OFDM achieves the lowest PAPR among the considered waveforms. Although the proposed DBU-OFDM does not completely attain the same performance as block-pilot DFT-s-OFDM in the low-PAPR region, it exhibits a significantly improved tail distribution. Specifically, due to the tail-aware design criterion adopted in the proposed framework, DBU-OFDM achieves performance close to that of block-pilot DFT-s-OFDM in the high-PAPR region.

It is further observed that the PAPR performance of DFT-s-OFDM degrades substantially when comb-type pilots are employed. This degradation is mainly caused by the puncturing effect introduced by comb-type pilot insertion on the DFT-spread signal, which destroys the single-carrier-like structure that gives rise to the low-PAPR characteristic of DFT-s-OFDM \cite{sahin2017dft}. Consequently, increasing the comb-pilot density results in more severe PAPR degradation.

Moreover, the PAPR performance under different modulation orders is also evaluated in Fig. \ref{fig:PAPR_all}. Although DBU-OFDM is trained under 16QAM, it maintains a CCDF tail behavior comparable to that of block-type DFT-s-OFDM when applied to 64QAM. This result further indicates that the proposed DBU-OFDM design generalizes well across modulation orders.

A more practical setting is also considered, where the transmitted signal is evaluated with an oversampling factor of 4. The corresponding CCDF curves of the shaped oversampled signals are shown in Fig. \ref{fig:PAPR_4times}. As expected, oversampling leads to a noticeable rightward shift of all CCDF curves, and block-type DFT-s-OFDM still achieves the lowest PAPR among all considered schemes. Since oversampling and pulse shaping are not explicitly incorporated into the training stage, DBU-OFDM exhibits an approximate performance loss of 0.7 dB relative to the block-type DFT-s-OFDM counterpart. However, in the practically relevant $10^{-3}$ to $10^{-4}$ CCDF region, it still provides more than 3 dB gain over conventional OFDM.

Overall, the proposed DBU-OFDM achieves PAPR performance close to that of block-pilot DFT-s-OFDM while retaining the comb-type pilot structure. This feature is particularly desirable in time-varying channels, where comb-type pilots are generally more effective than block-type pilots for channel tracking.

\subsection{Frequency-Diversity Enhancement}
In this subsection, we explain the communication gain of DBU-OFDM in frequency-selective fading, which mainly comes from the trainable mixing introduced within the data-subcarrier domain. In conventional OFDM, each data symbol is carried by a single subcarrier, and its reliability is therefore directly limited by the instantaneous fading condition of that subcarrier. In contrast, DBU-OFDM applies a unitary transformation only over the data subcarriers before transmission, such that each data symbol is distributed across multiple frequency components while the pilot and null subcarriers remain structurally protected. As a result, DBU-OFDM can exploit frequency-domain diversity without destroying the standard OFDM signal structure or the low-complexity frequency-domain equalization framework.  

To evaluate this effect, we consider communication over a frequency-selective fading channel.\footnote{A two-ray Rayleigh fading channel is considered. Specifically, the channel impulse response consists of two independent taps, each modeled as a zero-mean complex Gaussian random variable with unit variance, i.e., $h_i\sim\mathcal{CN}(0,1)$, $i\in\{1,2\}$.} According to Eq. \eqref{eqn:dbu_rx}, for the $m$-th OFDM symbol, the received signal on the data subcarriers can be written as
\begin{align}
\mathbf{Y}_{d,m}
=
\mathbf{\Lambda}_{d,m}\mathbf{U}_{\mathrm{data}}\mathbf{s}_{d,m}
+
\mathbf{W}_{d,m},
\end{align}
where $\mathbf{s}_{d,m}\in\mathbb{C}^{N_{\mathrm{data}}\times 1}$ denotes the transmitted data-symbol vector, $\mathbf{U}_{\mathrm{data}}\in\mathbb{C}^{N_{\mathrm{data}}\times N_{\mathrm{data}}}$ is the trainable unitary block acting on the data subcarriers, $\mathbf{\Lambda}_{d,m}$ is the diagonal channel matrix over the data subcarriers, and $\mathbf{W}_{d,m}\sim\mathcal{CN}(\mathbf{0},\sigma_0^2\mathbf{I})$ is the additive noise vector. Note that $\mathbf{\Lambda}_{d,m}$ is assumed to remain constant within the considered coherent processing block. 

At the receiver, conventional one-tap minimum mean square error (MMSE) equalization is adopted. Let $\mathbf{G}_{d,m}$ denotes the equalizer matrix, the equalized transformed-symbol vector is then expressed as $\hat{\bar{\mathbf{s}}}_{d,m}=\mathbf{G}_{d,m}\mathbf{Y}_{d,m}$, and the recovered data-symbol estimate is obtained through the inverse unitary transformation as 
\begin{align}
\hat{\mathbf{s}}_{d,m}
=
\mathbf{U}_{\mathrm{data}}^{H}\hat{\bar{\mathbf{s}}}_{d,m}.
\end{align}
Substituting $\mathbf{Y}_{d,m}$ into the above equations yields
\begin{align}
\hat{\mathbf{s}}_{d,m}
=
\mathbf{U}_{\mathrm{data}}^{H}\mathbf{G}_{d,m}\mathbf{\Lambda}_{d,m}\mathbf{U}_{\mathrm{data}}\mathbf{s}_{d,m}
+
\mathbf{U}_{\mathrm{data}}^{H}\mathbf{G}_{d,m}\mathbf{W}_{d,m}.\label{eqn:coupledChannel}
\end{align}
Eq. \eqref{eqn:coupledChannel} shows that, after equalization and inverse transformation, the effective post-equalization channel is no longer governed by a single scalar subcarrier coefficient, but by the coupled operator $\mathbf{U}_{\mathrm{data}}^{H}\mathbf{G}_{d,m}\mathbf{\Lambda}_{d,m}\mathbf{U}_{\mathrm{data}}$. Therefore, different from conventional OFDM, where a deep fade on one subcarrier directly causes a severe reliability loss for the symbol carried on that subcarrier, DBU-OFDM spreads the information of each data symbol across multiple subcarriers within the transformed data block. As a consequence, the final symbol reliability depends on the aggregate channel condition over the involved subcarriers, which constitutes the main source of frequency-diversity enhancement in the proposed waveform. 

To optimize this diversity effect from an end-to-end communication perspective, $\mathbf{U}_{\text{data}}$ is trained using a bit-wise supervised objective. After MMSE equalization and inverse transformation, soft bit metrics are constructed from the recovered constellation samples through the corresponding differentiable demapping procedure. These soft outputs are then compared with the transmitted bits via the binary cross-entropy (BCE) loss. Let $N_{\mathrm{bit}}$ denote the total number of transmitted bits in one training batch. The communication loss is defined as
\begin{align}
\mathcal{L}_{\mathrm{comm}}
=
-\frac{1}{N_{\mathrm{bit}}}
\sum_{i=1}^{N_{\mathrm{bit}}}
\left(
d_i\log_2 \hat d_i
+
(1-d_i)\log_2(1-\hat d_i)
\right),
\end{align}
where $d_i\in\{0,1\}$ denotes the $i$-th transmitted bit and $\hat d_i$ denotes the predicted probability that the $i$-th bit is one. This objective provides a fully differentiable criterion for optimizing the trainable unitary transformation under frequency-selective fading. 

\begin{figure}
	\centering
	\includegraphics[width=0.95\linewidth]{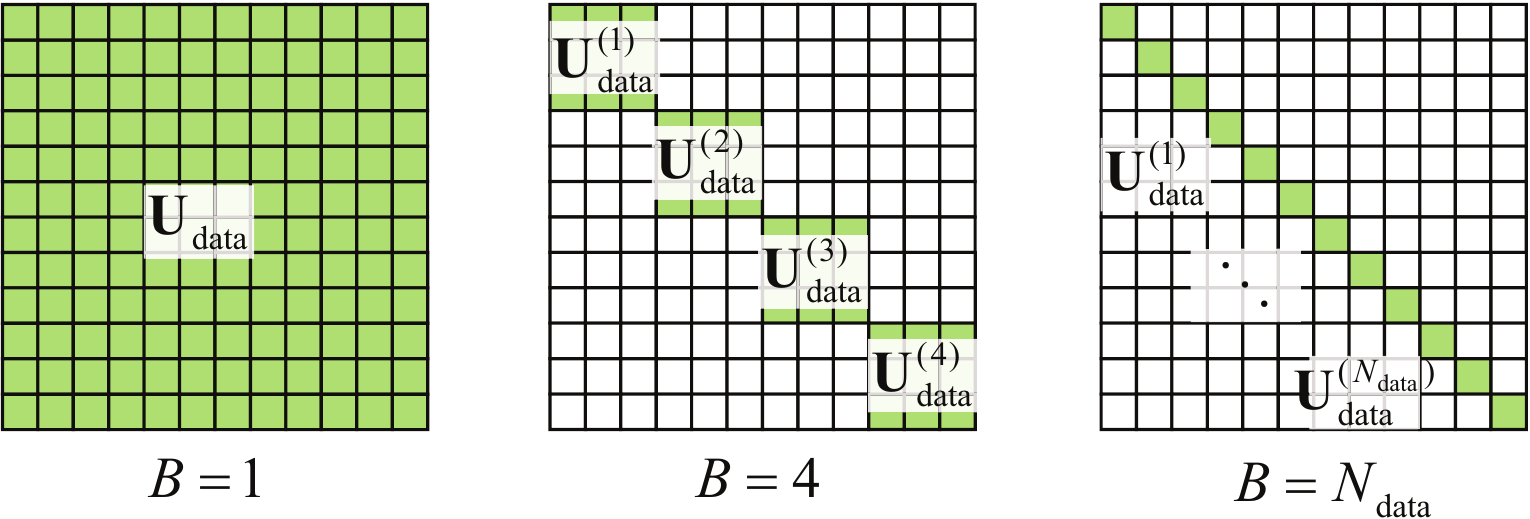}
	\caption{Block-wise structure of $\mathbf{U}_{\mathrm{data}}$ with cases $B=1$, $B=4$, and $B=N_{\mathrm{data}}$, respectively.}
    \vspace{-10pt}
	\label{fig:blockUdata}
\end{figure}

Moreover, in addition to the fully coupled design, we further introduce a block-wise variant of $\mathbf{U}_{\mathrm{data}}$ to explicitly control the frequency-mixing range. Specifically, the $N_{\mathrm{data}}$ data subcarriers are further divided into $B$ groups, as illustrated in Fig. \ref{fig:blockUdata}, and the trainable unitary matrix is further constrained to the block-diagonal form as
\begin{align}
\mathbf{U}_{\mathrm{data}}
=
\mathrm{blkdiag}
\!\left(
\mathbf{U}_{\mathrm{data}}^{(1)},
\mathbf{U}_{\mathrm{data}}^{(2)},
\ldots,
\mathbf{U}_{\mathrm{data}}^{(B)}
\right),
\end{align}
where $\mathbf{U}_{\mathrm{data}}^{(b)}\in\mathbb{C}^{N_b\times N_b}$ is the unitary transform associated with the $b$-th subcarrier group, and $\sum_{b=1}^{B} N_b = N_{\mathrm{data}}$. In this work, the data subcarriers are partitioned according to their original frequency order for simplicity, i.e., they are sequentially divided into contiguous groups. 
Such a grouping directly controls the mixing span in the frequency domain, while more sophisticated grouping strategies are left for future investigation.


\begin{figure}
	\centering
	
	\subfloat[$N=64$]{
		\includegraphics[width=0.8\linewidth]{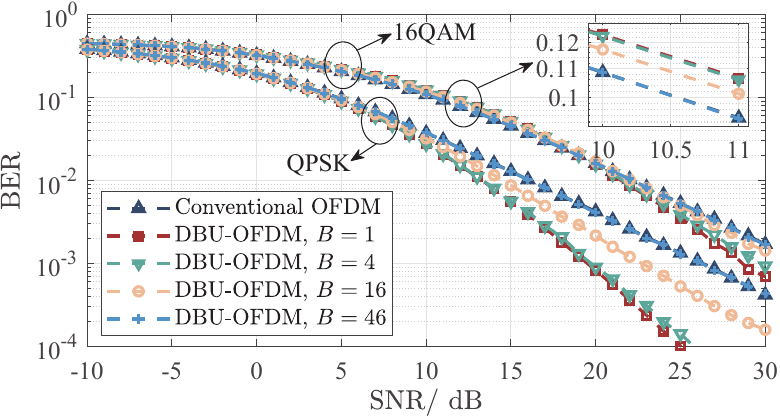}\label{fig:BERN64}
	}\\
	\vspace{-10pt}
	\subfloat[$N=256$]{
		\includegraphics[width=0.8\linewidth]{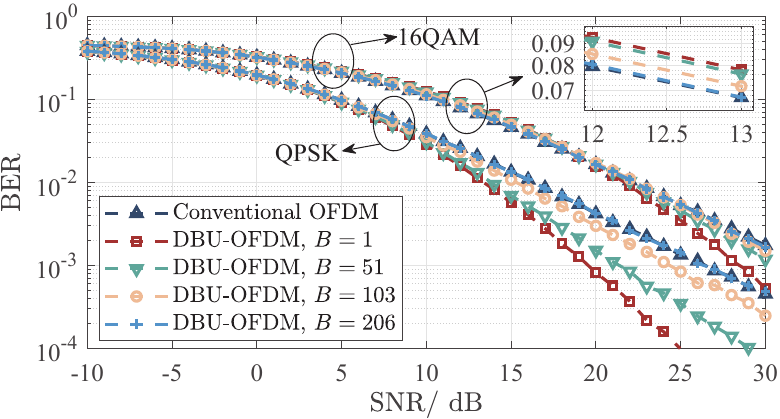}\label{fig:BERN256}
	}
	\caption{BER performance of conventional OFDM and DBU-OFDM under the considered frequency-selective fading channel for different block numbers $B$. Results are shown for $N=64$ and $N=256$ under QPSK and 16QAM.}
        \vspace{-10pt}
	\label{fig:BER1}
\end{figure}

Fig. \ref{fig:BER1} shows the BER performance under the considered frequency-selective fading channel. When $B=1$, all data subcarriers are jointly mixed by a single $\mathbf{U}_{\mathrm{data}}$, corresponding to the largest frequency-diversity range. In this case, DBU-OFDM achieves a lower BER than conventional OFDM in the high-SNR region for both $N=64$ and $N=256$. As $B$ increases, the mixing is restricted to smaller sub-blocks, and the diversity gain gradually decreases. In the limiting case $B=N_{\mathrm{data}}$, $\mathbf{U}_{\mathrm{data}}$ reduces to a diagonal phase matrix as shown in Fig. \ref{fig:blockUdata}, and the resulting communication behavior approaches that of conventional OFDM as shown in Fig. \ref{fig:BER1}.

A further observation is that DBU-OFDM consistently improves the BER of QPSK, whereas for 16QAM it exhibits a slightly higher BER than conventional OFDM in the medium-SNR regime. This is because, after one-tap MMSE equalization and inverse transformation, the effective channel is governed by the coupled operator $\mathbf{U}_{\mathrm{data}}^{H}\mathbf{G}_{d,m}\mathbf{\Lambda}_{d,m}\mathbf{U}_{\mathrm{data}}$ rather than an ideal per-subcarrier scalar channel. Thus, frequency-domain diversity and residual mixing penalty coexist in this regime. As the SNR increases, the latter becomes less dominant, while the diversity gain becomes more evident, such that DBU-OFDM eventually outperforms conventional OFDM in the high-SNR region.

\begin{figure}
	\centering
	
	\subfloat[$N=64$]{
		\includegraphics[width=0.8\linewidth]{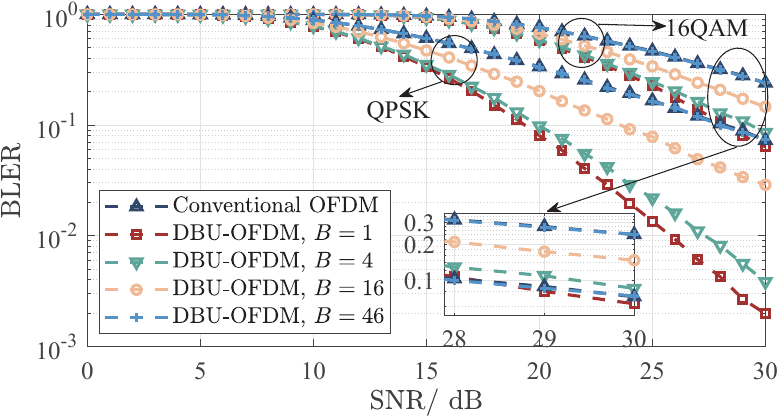}\label{fig:BLERN64}
	}\\
	\vspace{-10pt}
	\subfloat[$N=256$]{
		\includegraphics[width=0.8\linewidth]{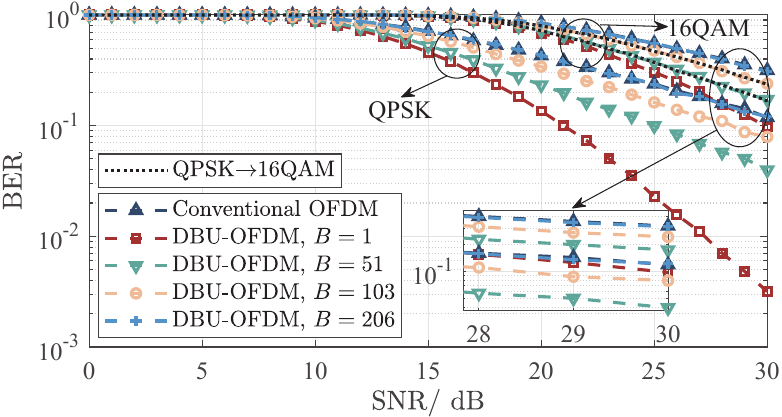}\label{fig:BLERN256}
	}
	\caption{BLER performance of conventional OFDM and DBU-OFDM under the considered frequency-selective fading channel for different block numbers $B$. The label “QPSK$\to$16QAM” indicates that $\mathbf{U}_{\mathrm{data}}$ is trained under QPSK and directly tested under 16QAM. Each bit block contains $N_{\mathrm{data}}\times M\times \log_2(O)$ bits, where $O$ denotes the modulation order.}
        \vspace{-10pt}
	\label{fig:BLER}
\end{figure}

The block error rate (BLER) results in Fig. \ref{fig:BLER} provide even clearer evidence of the benefit of DBU-OFDM. Over the entire SNR range, DBU-OFDM consistently outperforms conventional OFDM, and the gain becomes stronger as the number of blocks decreases. This behavior is fully consistent with the proposed mechanism. By spreading each data symbol across multiple subcarriers, DBU-OFDM reduces the sensitivity of a code block to a few deeply faded subcarriers, thereby significantly lowering the probability of block failure. Therefore, although the BER gain may not always be dominant in the medium-SNR regime, the BLER improvement remains clear and consistent.

Moreover, the block-wise results directly verify that the communication gain is governed by the frequency-mixing range introduced by $\mathbf{U}_{\mathrm{data}}$. A smaller number of blocks $B$ enables broader subcarrier mixing and stronger diversity, whereas a larger number of blocks restricts the mixing range and correspondingly reduces the gain.

Finally, Fig. \ref{fig:BLERN256} shows the BLER performance obtained by directly applying the $\mathbf{U}_{\mathrm{data}}$ trained under QPSK to 16QAM. The results suggest that the main benefit of DBU-OFDM is primarily rooted in the channel-dependent frequency-diversity enhancement induced by $\mathbf{U}_{\mathrm{data}}$, and that the learned transformation $\mathbf{U}_\text{data}$ retains a certain degree of transferability across different modulation orders.

\subsection{Direct Integrated Sensing Capability}
Beyond communication enhancement, the proposed DBU-OFDM waveform can also directly support sensing within the same OFDM framework.

Specifically, let $\bar{\mathbf{s}}_m=\mathbf{U}\mathbf{s}_m$ denote the effective frequency-domain symbol vector of the $m$-th OFDM symbol. Consider a sensing scene composed of $L$ dominant propagation paths, where the $\ell$-th path is characterized by a complex reflection coefficient $\beta_\ell$, a propagation delay $\tau_\ell$, and a Doppler frequency $\nu_\ell$. After CP removal and DFT demodulation, the received signal on the $n$-th subcarrier of the $m$-th OFDM symbol can be modeled as
\begin{align}
	Y_{n,m}
	=
	\sum_{\ell=1}^{L}
	\beta_\ell \bar{S}_{n,m}
	e^{-j2\pi (n-1)\Delta f \tau_\ell}
	e^{j2\pi (m-1)T_o \nu_\ell}
	+
	W_{n,m},
\end{align}
where $\bar{S}_{n,m}$ denotes the $n$-th entry of $\bar{\mathbf{s}}_m$, $\Delta f$ is the subcarrier spacing, and $W_{n,m}$ is the additive noise. 

Since the effective transmitted symbols $\bar{S}_{n,m}$ are known at the receiver, an initial sensing observation can be constructed through element-wise matched demodulation as
\begin{align}
Z_{n,m}&=\frac{Y_{n,m}\bar{S}_{n,m}^{\dagger}}{|\bar{S}_{n,m}|}\nonumber\\
&=\sum_{\ell=1}^{L}
\beta_\ell
|\bar{S}_{n,m}|e^{-j2\pi (n-1)\Delta f \tau_\ell}
e^{j2\pi (m-1)T_o \nu_\ell}
+ W_{n,m}.
\end{align}
where $\bar{S}_{n,m}^\dagger W_{n,m}/|\bar{S}_{n,m}|=e^{-j2\pi\phi_{n,m}}W_{n,m}$, and $\phi_{n,m}$ denotes the phase introduced by $\bar{S}_{n,m}^\dagger$. Therefore, the power of $W_{n,m}$ becomes $\sigma_0^2e^{-j2\pi\phi_{n,m}}e^{j2\pi\phi_{n,m}}=\sigma_0^2$, indicating that the phase $\phi_{n,m}$ does not affect the noise power.

Stacking $Z_{n,m}$ over the OFDM-symbol and subcarrier dimensions gives an observation matrix $\mathbf{Z}\in\mathbb{C}^{M\times N}$, whose $(m,n)$-th entry is $Z_{n,m}$. We further define the delay steering vector and Doppler steering vector as
\begin{align}
\mathbf{a}(\tau)=
\left[
1,\,
e^{-j2\pi \Delta f\tau},\,
\ldots,\,
e^{-j2\pi (N-1)\Delta f\tau}
\right]^T,
\end{align}
and
\begin{align}
\mathbf{b}(\nu)=
\left[
1,\,
e^{j2\pi T_o\nu},\,
\ldots,\,
e^{j2\pi (M-1)T_o\nu}
\right]^T,
\end{align}
respectively. Based on these steering vectors, the two-dimensional delay-Doppler correlation metric is defined as
\begin{align}
\mathcal{L}(\tau,\nu)
=
\left|
\mathbf{b}^H(\nu)\mathbf{Z}\mathbf{a}(\tau)
\right|^2.
\end{align}
When the candidate pair $(\tau,\nu)$ matches an actual path, the phase rotations induced by delay and Doppler are effectively compensated, such that the signal components across subcarriers and symbols are coherently combined and a pronounced peak is produced in the delay-Doppler map. In contrast, when the candidate parameters are mismatched, the phase terms cannot be properly aligned, and the resulting summation is largely suppressed due to destructive interference.

Generally, the path parameters are obtained through hard peak selection as 
\begin{align}
(\hat{\tau},\hat{\nu})
=
\arg\max_{\tau,\nu}\mathcal{L}(\tau,\nu).
\end{align}
However, the above hard-argmax operation is non-differentiable and therefore cannot be directly incorporated into end-to-end optimization of the trainable unitary matrix. To address this issue, we adopt a differentiable soft-max-based estimator. Specifically, for a discrete delay-Doppler search grid $(\tau_i,\nu_j)$, we define
\begin{align}
p_{i,j}
=
\frac{\exp\!\big(\gamma \mathcal{L}(\tau_i,\nu_j)\big)}
{\sum_{i',j'} \exp\!\big(\gamma \mathcal{L}(\tau_{i'},\nu_{j'})\big)},
\end{align}
where $\gamma$ is a temperature parameter controlling the sharpness of the resulting distribution. The delay and Doppler estimates are then computed as the corresponding expectations over the search grid
\begin{align}
\hat{\tau}=\sum_{i,j}\tau_i\,p_{i,j},
\qquad
\hat{\nu}=\sum_{i,j}\nu_j\,p_{i,j}.
\end{align}
This formulation preserves the likelihood-based estimation structure while enabling gradient back-propagation through the sensing module. Moreover, it can exhibit sub-grid estimation behavior when the likelihood mass is concentrated around a local peak.

To further support multipath sensing, an iterative successive interference cancellation (SIC) procedure is employed. Let $\mathbf{Z}^{(q)}$ denote the residual observation matrix at the $q$-th iteration. After obtaining the strongest path parameters $(\hat{\tau}_q,\hat{\nu}_q)$ from the above likelihood search, the associated complex coefficient can be estimated by
\begin{align}
\hat{\beta}_q
=
\frac{\mathbf{b}^H(\hat{\nu}_q)\mathbf{Z}^{(q)}\mathbf{a}(\hat{\tau}_q)}
{\|\mathbf{b}(\hat{\nu}_q)\|_2^2 \, \|\mathbf{a}(\hat{\tau}_q)\|_2^2}.
\end{align}
The corresponding path contribution is then removed from the residual matrix according to
\begin{align}
\mathbf{Z}^{(q+1)}
=
\mathbf{Z}^{(q)}
-
\hat{\beta}_q\,\mathbf{b}(\hat{\nu}_q)\mathbf{a}^H(\hat{\tau}_q).
\end{align}
This procedure is repeated until the pre-specified number of paths is reached. In this way, dominant paths can be successively extracted and canceled, thereby alleviating mutual masking among echoes and improving the resolvability of weaker paths. Thus, after $L$ SIC iterations, the estimated delay-Doppler parameters of the extracted dominant paths are given by $\{(\hat{\tau}_q,\hat{\nu}_q)\}_{q=1}^{L}$. Accordingly, the corresponding range and velocity estimates can be written as
\begin{align}
\hat{r}_q=\frac{c\hat{\tau}_q}{2},\quad
\hat{v}_q=\frac{\lambda_c \hat{\nu}_q}{2},\quad q=1,\ldots,L,
\end{align}
where $c$ is the speed of light and $\lambda_c$ denotes the carrier wavelength. Let $\{(r_q,v_q)\}_{q=1}^{L}$ denote the ground-truth range and velocity of the $L$ dominant paths, indexed consistently with the SIC extraction order. Then, a sensing-oriented training loss can be defined as the average mean square error (MSE) of the extracted multipath estimates
\begin{equation}
	\mathcal{L}_{\mathrm{sens}}
	=
	\frac{1}{2L}\sum_{q=1}^{L}
	\left[
	(\hat{r}_q-r_q)^2+(\hat{v}_q-v_q)^2
	\right].
\end{equation}

\begin{figure}
	\centering
	
	\subfloat[Range estimation]{
		\includegraphics[width=0.85\linewidth]{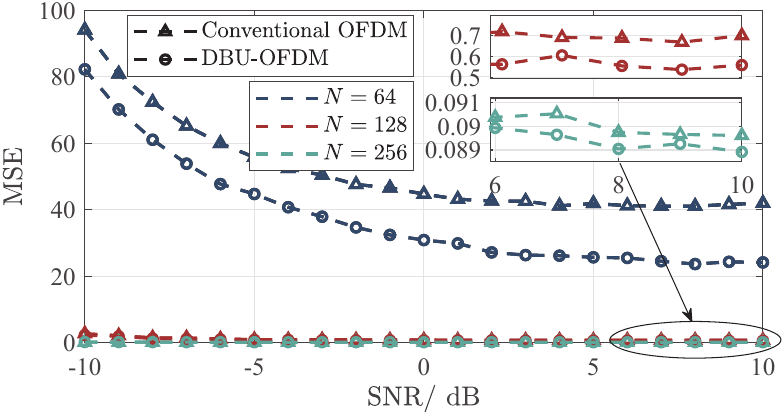}\label{fig:MSEest_range}
	}\\
	\vspace{-10pt}
	\subfloat[Velocity estimation]{
		\includegraphics[width=0.85\linewidth]{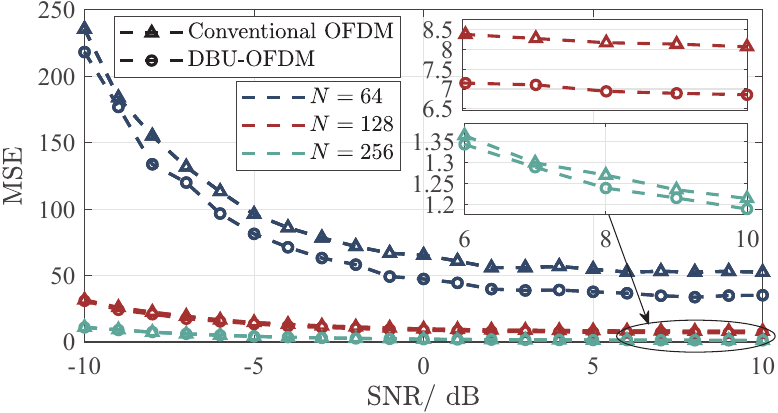}\label{fig:MSEest_velo}
	}
	\caption{Sensing performance comparison between conventional OFDM and DBU-OFDM under the three parameter configurations $N=\{64,128,256\}$ and $L = 3$.}
        \vspace{-10pt}
	\label{fig:MSE_delay_doppler}
\end{figure}

We then evaluate the sensing performance of proposed DBU-OFDM. Fig. \ref{fig:MSE_delay_doppler} compares the sensing performance of conventional OFDM and the proposed DBU-OFDM in terms of range and velocity estimation MSE under the three parameter settings listed in Table \ref{tab:ofdm_parameters}. As shown in Fig. \ref{fig:MSEest_range}, DBU-OFDM consistently achieves lower range estimation MSE than conventional OFDM across the entire SNR range for all considered values of $N$. The gain is most pronounced for $N=64$, particularly in the medium-to-high SNR regime, indicating that the proposed waveform is especially beneficial when the sensing observation dimension is limited. In this case, the learned block-unitary transformation improves the robustness of delay estimation and enhances the separability of multipath components.

As $N$ increases from 64 to 128 and 256, the performance gap gradually decreases. This trend suggests that, with a larger frequency-domain observation aperture, conventional OFDM already provides improved delay resolution and more stable parameter estimation, thereby leaving less room for further improvement. However, DBU-OFDM maintains a consistent sensing advantage across all considered settings. 

A similar trend is observed in Fig. \ref{fig:MSEest_velo} for velocity estimation, where DBU-OFDM also yields lower MSE than conventional OFDM for all configurations, while the relative gain becomes smaller as the number of subcarriers increases.

Overall, Fig. \ref{fig:MSE_delay_doppler} indicates that DBU-OFDM improves both range and velocity sensing performance within the same OFDM sensing framework. The improvement is particularly evident in small-subcarrier settings, where sensing resources are limited and multipath effects are more difficult to resolve, whereas the relative advantage becomes smaller once the observation aperture is sufficiently large.

\section{Prototype Validation}\label{sec:V}
In this section, we first conduct an initial over-the-air validation of conventional OFDM and DBU-OFDM on a USRP-based platform, and then present a detailed FPGA design and performance evaluation.

\subsection{USRP-Based Validation}

\begin{figure}
	\centering
	\includegraphics[width=0.8\linewidth]{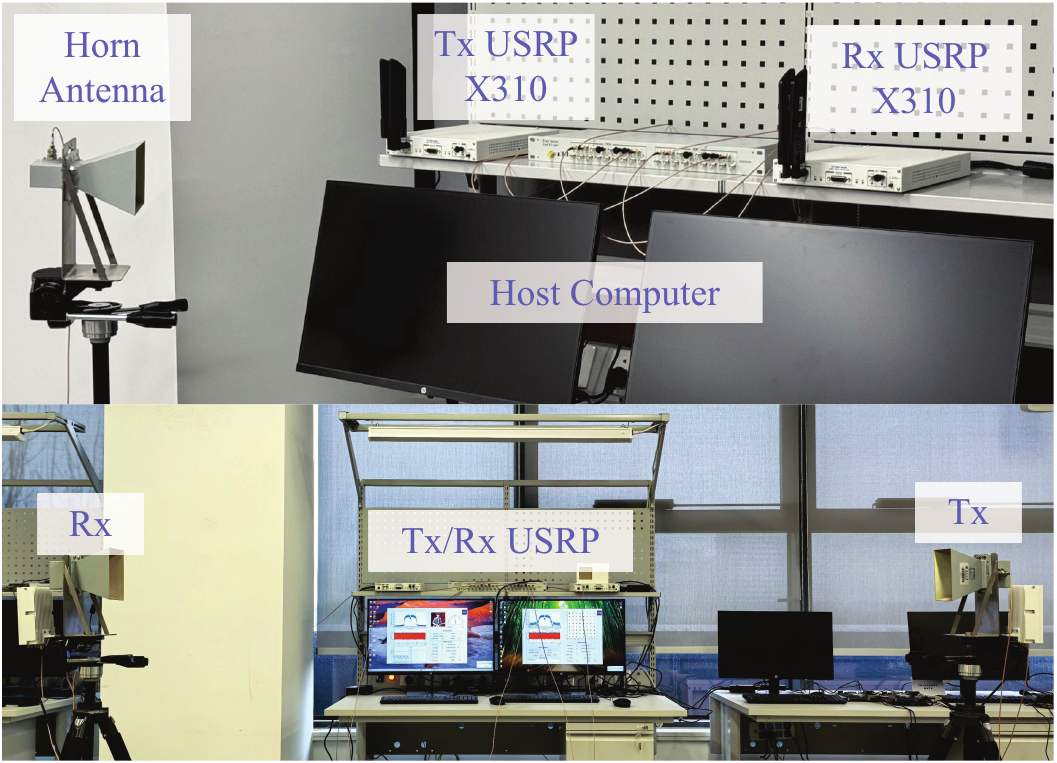}
	\caption{USRP-based over-the-air experimental setup for validating conventional OFDM and DBU-OFDM. }
        \vspace{-10pt}
	\label{fig:envsUSRP}
\end{figure}

The USRP-based experimental setup is shown in Fig. \ref{fig:envsUSRP}. Two USRP X310 devices equipped with UBX-160 daughterboards are employed. Together with two horn antennas, the signals are upconverted to 3.6 GHz for wireless transmission. The transmitter and receiver are controlled by separate host computers. The Tx-Rx distance is 2.5 m with aligned beam directions, and one-tap MMSE equalization is adopted at the receiver for both schemes.

\begin{figure}
	\centering
	\subfloat[OFDM, EVM=1.2823\%]{
		\includegraphics[width=0.46\linewidth]{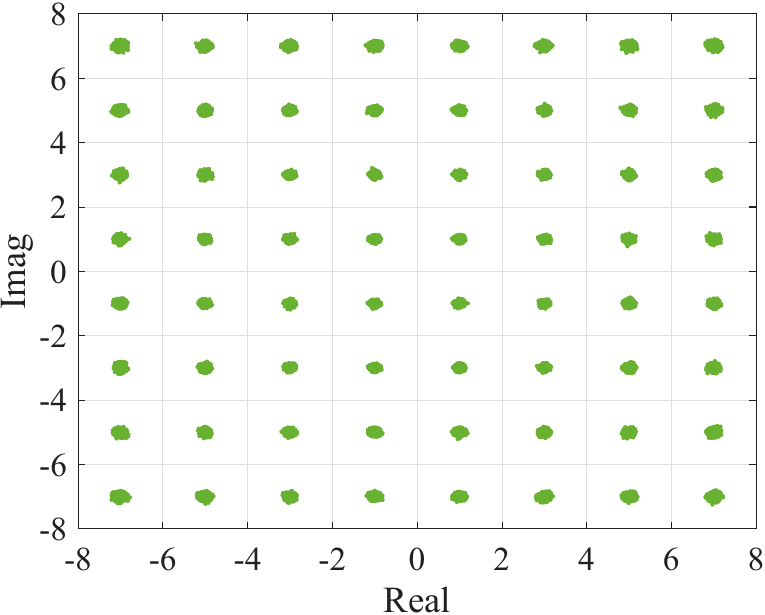}\label{fig:USRPcons_org}
	}
	\subfloat[DBU-OFDM, EVM=1.2780\%]{
		\includegraphics[width=0.46\linewidth]{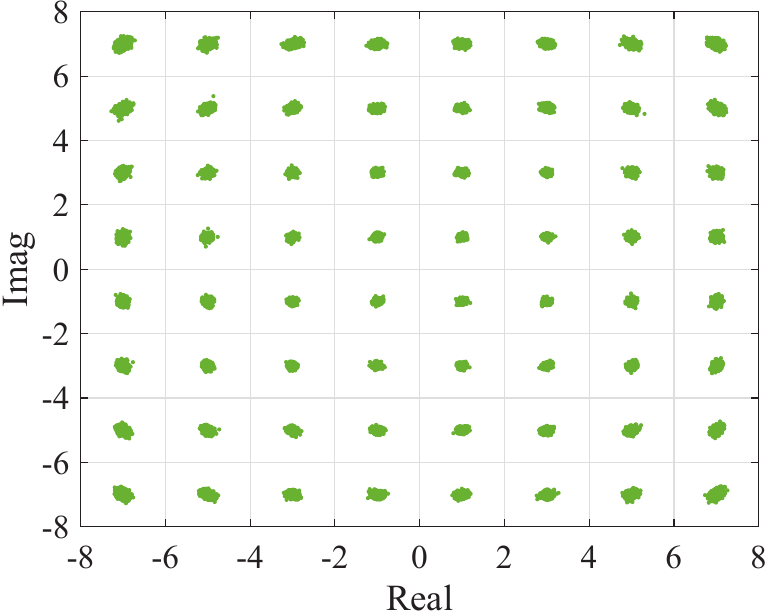}\label{fig:USRPcons_dbu}
	}
	\caption{Received constellations in the USRP-based over-the-air experiment with $N = 256$ and 64QAM.}
        \vspace{-10pt}
	\label{fig:USRPcons}
\end{figure}

\begin{figure}[!t]
	\centering
	\subfloat[]{
		\includegraphics[width=0.8\linewidth]{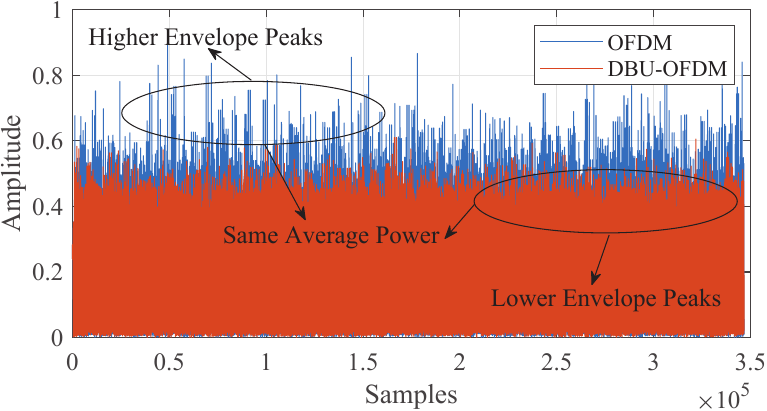}\label{fig:USRPPAPRabs}
	}\\
	\vspace{-10pt}
	\subfloat[]{
		\includegraphics[width=0.8\linewidth]{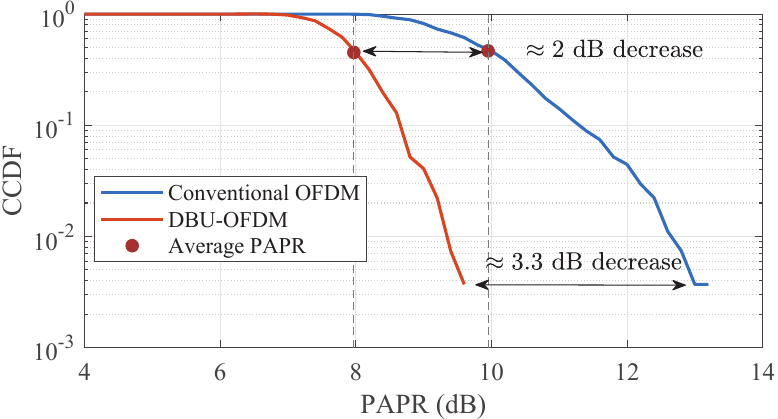}\label{fig:USRPCCDF}
	}
	\caption{Experimental PAPR comparison between conventional OFDM and DBU-OFDM in the USRP-based validation. (a) Absolute envelope of the transmitted time-domain waveforms. (b) Measured CCDF of PAPR.}
        \vspace{-10pt}
\end{figure}

Fig. \ref{fig:USRPcons} illustrates the measured communication performance of conventional OFDM and DBU-OFDM, where $N=256$ and 64QAM is adopted. Note that in the practical over-the-air experiments, the OFDM signal maintains satisfactory communication performance under the current experimental setup, with an error vector magnitude (EVM) of only $1.2823\%$. Although DBU-OFDM is trained solely for PAPR reduction, it still preserves almost identical communication performance to that of conventional OFDM, achieving an EVM of only $1.2780\%$ for the 64QAM modulation. Meanwhile, a slight constellation rotation can be observed in DBU-OFDM, which might be caused by residual phase errors, including phase noise, carrier frequency offset (CFO), and common phase error (CPE). However, DBU-OFDM still exhibits clearly distinguishable 64 symbol states, which further demonstrates its robustness against practical non-ideal impairments.

Fig. \ref{fig:USRPPAPRabs} shows the absolute envelope of the transmitted time-domain waveforms for OFDM and DBU-OFDM over a total of 270 OFDM symbols. Note that under the same average transmit power, DBU-OFDM, which is designed with PAPR reduction as the objective, exhibits a significantly lower and smoother envelope. This indicates that DBU-OFDM can effectively reduce the peak power, and thus lower the PAPR, while preserving the communication capability.

Moreover, statistical analysis of the transmitted signals was conducted, and the corresponding complementary CCDF curves of the PAPR are presented in Fig. \ref{fig:USRPCCDF}. It is observed that, compared with conventional OFDM, DBU-OFDM achieves an average PAPR reduction of approximately 2 dB, while the reduction in the tail region reaches about 3.3 dB. These results further verify the effectiveness of the proposed waveform.

\subsection{FPGA Implementation and Efficiency Evaluation}

After the over-the-air validation on the USRP platform, we further implement the core block-unitary transform of the proposed DBU-OFDM on FPGA. It should be emphasized that the objective of the hardware design is not to reconstruct the entire OFDM baseband chain, but rather to efficiently map $\mathbf{U}_{\mathrm{data}}$ onto a pipelined and cascaded FPGA datapath.

According to Section \ref{sec:ii-c}, $\mathbf{U}_{\mathrm{data}}^{(k)}$ denotes the $k$-th Householder transformation matrix. To remain consistent with the definition of $\mathbf{v}_k \in \mathbb{C}^{1\times N_{\mathrm{data}}}$, we further define the corresponding normalized column vector as
\begin{align}
\mathbf{u}_k
\triangleq
\frac{\mathbf{v}_k^H}{\sqrt{\mathbf{v}_k\mathbf{v}_k^H}}
\in\mathbb{C}^{N_{\mathrm{data}}\times 1},
\end{align}
which yields
\begin{align}
\mathbf{U}_{\mathrm{data}}^{(k)}
=
\mathbf{I}_{N_{\mathrm{data}}}
-
2\mathbf{u}_k\mathbf{u}_k^H.
\end{align}
Accordingly, the implementation of $\mathbf{U}_{\mathrm{data}}$ can be naturally decomposed into a cascaded structure consisting of multiple Householder modules followed by one diagonal phase-multiplication module $\mathbf{D}$, where the Householder chain captures cross-element coupling, whereas $\mathbf{D}$ only corresponds to element-wise complex phase rotation.

A direct realization of $\mathbf{U}_{\mathrm{data}}$ through matrix products would require repeated matrix-vector multiplications, which is unfavorable for regular pipelined processing and on-chip storage on FPGA. To address this issue, each Householder transformation is rewritten in recursive form. For the input vector $\mathbf{x}_{k-1}\in\mathbb{C}^{N_{\mathrm{data}}\times 1}$ of the $k$-th Householder stage, we have
\begin{align}
\mathbf{x}_k=\mathbf{x}_{k-1}-2\mathbf{u}_k\left(\mathbf{u}_k^H\mathbf{x}_{k-1}\right),
\quad k=1,2,\ldots,K.
\end{align}
Therefore, each Householder operation can be uniformly decomposed into three steps: inner product, scaling, and update. Specifically, the complex scalar $\mathbf{u}_k^H\mathbf{x}_{k-1}$ is first computed, then multiplied by the vector $\mathbf{u}_k$ to form the correction term, and finally subtracted twice from the input vector $\mathbf{x}_{k-1}$ to obtain $\mathbf{x}_k$. This formulation avoids explicit construction of the full matrix and is therefore more suitable for a regular cascaded pipeline architecture.

Meanwhile, since each Householder matrix is both Hermitian and unitary, and the inverse of $\mathbf{D}$ only corresponds to element-wise conjugate phase rotation, the receiver-side inverse transform and the transmitter-side forward transform can reuse the same basic computational units at the hardware level, with only the Householder cascade order reversed between the two sides, i.e.,
\begin{align}
\mathbf{U}_{\mathrm{data}}^H
=
\left(\mathbf{D}\prod_{i=1}^{K}\mathbf{U}_{\mathrm{data}}^{(i)}\right)^H
=
\prod_{i=K}^{1}\left(\mathbf{U}_{\mathrm{data}}^{(i)}\right)^H\mathbf{D}^H.
\end{align}
In other words, the proposed architecture does not require two different hardware modules for the transmitter and receiver. Instead, it enables bidirectional reuse through a unified parameterized computational kernel, which is beneficial for reducing the overall hardware complexity.

On this basis, we further merge two adjacent Householder stages to reduce inter-stage buffering and intermediate data movement. 

Taking two consecutive stages as an example, let the input be $\mathbf{x}_0$ and output be $\mathbf{x}_2$, we have
\begin{align}
\mathbf{x}_1
&=
\left(\mathbf{I}_{N_{\mathrm{data}}}-2\mathbf{u}_1\mathbf{u}_1^H\right)\mathbf{x}_0,\nonumber\\
\mathbf{x}_2
&=
\left(\mathbf{I}_{N_{\mathrm{data}}}-2\mathbf{u}_2\mathbf{u}_2^H\right)\mathbf{x}_1.
\end{align}
Define three complex scalars as
\begin{align}
\alpha_1=\mathbf{u}_1^H\mathbf{x}_0,\quad
\alpha_2=\mathbf{u}_2^H\mathbf{x}_0,\quad
\rho=\mathbf{u}_2^H\mathbf{u}_1,
\end{align}
and the joint output of the two stages can be written as
\begin{align}
\mathbf{x}_2
=
\mathbf{x}_0
-
\left[
2\alpha_1\mathbf{u}_1
+
2\left(\alpha_2-2\alpha_1\rho\right)\mathbf{u}_2
\right].
\end{align}
This result shows that the merged two-stage output $\mathbf{x}_2$ no longer depends explicitly on the intermediate vector $\mathbf{x}_1$, but can instead be computed directly from $\mathbf{x}_0$ and the scalars $\alpha_1$, $\alpha_2$, and $\rho$. Therefore, the original two serial Householder stages can be restructured as a single merged processing module.
\begin{table}
    \centering
    \caption{Hardware Implementation Comparison}
    \label{tab:hardware_comparison}
    \footnotesize
    \setlength{\tabcolsep}{3pt}
    \resizebox{\columnwidth}{!}{%
    \begin{tabular}{l c c c c c c c c}
        \toprule
        \textbf{Platform} & \textbf{\begin{tabular}[c]{@{}c@{}}Data\\ Width\\ (bit)\end{tabular}} & \textbf{\begin{tabular}[c]{@{}c@{}}Latency\\ ($\mu$s)\end{tabular}} & \textbf{\begin{tabular}[c]{@{}c@{}}Rate\\ (MS/s)\end{tabular}} & \textbf{BRAM} & \textbf{DSP} & \textbf{FF} & \textbf{LUT} & \textbf{\begin{tabular}[c]{@{}c@{}}Power\\ (W)\end{tabular}} \\
        \midrule
        Xilinx FFT IP\cite{xilinx2022fast}   & 16    & 8.38  & 30.5  & 5.5 & 9    & 3723  & 2068  & 0.4  \\
        \cite{zhou2023flexible}              & 16    & 1.4   & 182.8 & 2   & 28   & 2434  & 3308  & -    \\
        \cite{shi2025area}                   & 16    & 1.31  & 195.4 & 8   & 0    & 1941  & 10637 & 0.45 \\
        \cite{hazarika2023efficient}         & 12    & 2.61  & 200.0 & 1   & 0    & 4440  & 3467  & -    \\
        Proposed($K=4$)                      & Mixed & 2.06  & 200.0 & 0   & 48   & 894   & 2084  & 0.49 \\
        Proposed($K=32$)                     & Mixed & 16.48 & 200.0 & 0   & 384  & 6752  & 16272 & 0.60 \\
        Proposed($K=128$)                    & Mixed & 65.92 & 199.9 & 0   & 1536 & 27008 & 65088 & 0.70 \\
        \bottomrule
    \end{tabular}%
    }
\end{table}

\begin{figure}
	\centering
	\includegraphics[width=0.75\linewidth]{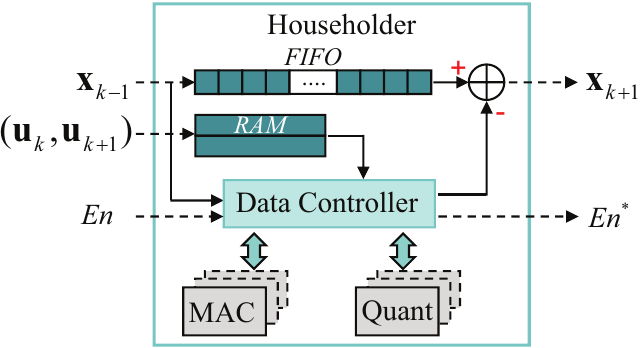}
	\caption{FPGA architecture of the merged Householder hardware module. The module comprises an input FIFO, parameter storage, complex multiply-accumulate (MAC) units, quantization units, and control logic. $\mathit{En}$ and $\mathit{En}^*$ denote the input-valid and output-valid signals, respectively. }
        \vspace{-10pt}
	\label{fig:householderFPGA}
\end{figure}

Based on the above merged two-stage formulation, a merged Householder hardware module is constructed, as illustrated in Fig. \ref{fig:householderFPGA}. For a network with $K$ Householder factors, the overall architecture can be realized by cascading $K/2$ merged Householder modules\footnote{A mixed fixed-point quantization strategy is used to balance performance and hardware cost. Specifically, $\mathbf{u}_k$, $\mathbf{x}$, and intermediate results are represented in $Q(12,10)$, $Q(10,6)$, and $Q(12,6)$ formats, respectively, where $Q(a,b)$ denotes $a$-bit fixed-point representation with $b$ fractional bits.}, followed by one element-wise complex multiplication module corresponding to $\mathbf{D}$.

Table \ref{tab:hardware_comparison} compares the synthesis results of the proposed implementation of $\mathbf{U}_{\text{data}}$ with several FFT-based baselines. Different from \cite{zhou2023flexible} and \cite{shi2025area}, whose high throughput is primarily enabled by multi-parallel inputs, the Xilinx FFT IP \cite{xilinx2022fast}, \cite{hazarika2023efficient}, and the proposed design all employ serial-input architectures that are more consistent with practical communication interfaces. Moreover, both \cite{hazarika2023efficient} and the proposed design are fully pipelined, and the reported latency corresponds to the pipeline fill latency, after which one output sample is produced per clock cycle. For $K=4$, the proposed design achieves $200$ mega-samples per second ($\mathrm{MS/s}$) with a latency of $2.06~\mu s$, while requiring no BRAM and only 894 FFs and 2084 LUTs. As $K$ increases from 4 to 32 and 128, the throughput remains essentially unchanged, whereas the latency and hardware resource usage increase with the number of stages. Although the output of the Householder module must be followed by an FFT stage in the subsequent computation according to the overall transceiver design in Section \ref{sec:ii-c}, the resulting overhead remains acceptable\footnote{For most tasks, including communication and sensing enhancement, $K=16\sim32$ is already sufficient; only a few tasks, such as PAPR reduction, require $K=128$ to achieve satisfactory performance.}. This confirms that the proposed architecture offers a tunable complexity-performance tradeoff and is suitable for practical hardware realization.

\section{Conclusions}\label{sec:conclusion}

In this paper, we proposed DBU-OFDM, a structure-preserving learning framework that introduces trainable waveform adaptation into OFDM without sacrificing its core engineering advantages. By confining learning to a block-unitary transform over data subcarriers and parameterizing it via recursive Householder reflections, the proposed design preserves strict unitarity, protects pilot and null resources, and retains the low-complexity single-tap equalization structure of conventional OFDM. Results showed that DBU-OFDM achieves PAPR tails close to block-pilot DFT-s-OFDM while preserving comb-type pilots, improves communication reliability in frequency-selective fading through frequency-domain diversity, and enhances direct sensing performance in range and velocity estimation. USRP and FPGA results further verified its practical viability, demonstrating real-transmission PAPR reduction, nearly unchanged communication quality, and efficient hardware realization. Overall, DBU-OFDM provides a practical middle ground between conventional model-based OFDM waveforms and unconstrained black-box neural transceivers for future communication and sensing systems.

\bibliography{Reference}
\end{document}